\documentclass[12pt]{article}
\usepackage[left=2cm, right=2cm, bottom=2cm ,top = 2cm, footskip=1cm]{geometry}
\usepackage[font=small,labelfont=bf]{caption}
\usepackage{xcolor}
\usepackage{setspace}
\usepackage{chngcntr}
\usepackage{array,multirow}
\usepackage{hyperref}
\hypersetup{colorlinks,allcolors=black}
\usepackage{setspace}

\usepackage[authoryear,sort&compress]{natbib}
\setcitestyle{open={(},close={)}}
\usepackage{enumitem}
\usepackage{graphicx}
\usepackage{amsmath}
\usepackage{titlesec}
\titleformat{\paragraph}
{\normalfont\normalsize\bfseries}{\theparagraph}{1em}{}
\titlespacing*{\paragraph}
{0pt}{3.25ex plus 1ex minus .2ex}{1.5ex plus .2ex}

\usepackage[section]{placeins}

\title{A library of quantitative markers of seizure severity}
\author{Sarah J. Gascoigne$^{1*}$, Leonard Waldmann$^{6}$, Gabrielle M. Schroeder$^1$,\\ Mariella Panagiotopoulou$^1$, Jess Blickwedel$^1$, Fahmida Chowdhury$^3$, Alison Cronie$^4$,\\ Beate Diehl$^3$, John S. Duncan$^3$, Jennifer Falconer$^4$, Ryan Faulder$^1$, Yu Guan$^7$,\\ Veronica Leach$^4$, Shona Livingstone$^4$, Christoforos Papasavvas$^1$, \\Rhys H. Thomas$^2$, Kevin Wilson$^5$, Peter N. Taylor$^{1,2,3}$, \\ Yujiang Wang$^{1,2,3**}$}

\begin{document}

\maketitle

\begin{enumerate}
\item{CNNP Lab (www.cnnp-lab.com), Interdisciplinary Computing and Complex BioSystems Group, School of Computing, Newcastle University, Newcastle upon Tyne, United Kingdom}
\item{Faculty of Medical Sciences, Newcastle University, United Kingdom}
\item{UCL Queen Square Institute of Neurology, Queen Square, London, United Kingdom}
\item{NHS Greater Glasgow and Clyde, Glasgow, United Kingdom}
\item{School of Mathematics, Statistics \& Physics, Newcastle University, United Kingdom}
\item{Technical University Munich, Munich, Germany}
\item{Department of Computer Science, University of Warwick, Warwick, United Kingdom}
\end{enumerate}

\begin{itemize}[label={}]
\item $^*$ Sarah Jane Gascoigne (ORCID ID: 0000-0003-1013-1875)\\
Urban Sciences Building, 
1 Science Square Newcastle upon Tyne, NE4 5TG\\
Tel: (+44) 191 208 4145 \hskip5em Email: S.Gascoigne@newcastle.ac.uk
\item $^{**}$ Yujiang Wang (ORCID ID: 0000-0002-4847-6273)\\
Urban Sciences Building, 
1 Science Square Newcastle upon Tyne, NE4 5TG\\
Tel: (+44) 191 208 4141  \hskip5em
Email: Yujiang.Wang@newcastle.ac.uk
\end{itemize}

\begin{center}
Leonard.Waldmann@gmx.de;
Gabrielle.Schroeder$^a$ M.Panagiotopoulou2$^a$;
Jessica.Blickwedel$^a$;  Fahmidaamin.Chowdhury$^b$;
Alison.Cronie$^c$; B.Diehl$^d$; J.Duncan$^d$;
Jennifer.Falconer$^c$; 
Ryan.Faulder2$^b$;
Yu.Guan@warwick.ac.uk; Veronica.Leach$^c$;
Shona.Livingstone$^c$; cpapasavva@gmail.com;
Rhys.Thomas$^a$; Kevin.Wilson$^a$;
Peter.Taylor$^a$

$^a$ @newcastle.ac.uk
$^b$ @nhs.net
$^c$ @ggc.scot.nhs.uk
$^d$ @ucl.ac.uk
\end{center}

\noindent{
We confirm that we have read the Journal’s position on issues involved in ethical publication and affirm that this report is consistent with those guidelines. \\
None of the authors has any conflict of interest to disclose.}
\newpage

\begin{doublespace}

\section{Abstract}
\textbf{Objective:} Understanding fluctuations in seizure severity within individuals is important for determining treatment outcomes and responses to therapy, as well as assessing novel treatments for epilepsy. Current methods for grading seizure severity rely on qualitative interpretations from patients and clinicians. Quantitative measures of seizure severity would complement existing approaches, for electroencephalographic (EEG) monitoring, outcome monitoring, and seizure prediction. Therefore, we developed a library of quantitative EEG markers that assess the spread and intensity of abnormal electrical activity during and after seizures. 

\textbf{Methods:} We analysed intracranial EEG (iEEG) recordings of 1009 seizures from 63 patients. For each seizure we computed 16 markers of seizure severity that capture the signal magnitude, spread, duration, and post-ictal suppression of seizures. 

\textbf{Results:} Quantitative EEG markers of seizure severity distinguished focal vs. subclinical seizures across patients. In individual patients 53$\%$ had a moderate to large difference (ranksum $r > 0.3$, $p<0.05$) between focal and subclinical seizures in three or more markers. Circadian and longer-term changes in severity were found for the majority of patients. 

\textbf{Significance:} We demonstrate the feasibility of using quantitative iEEG markers to measure seizure severity. Our quantitative markers distinguish between seizure types and are therefore sensitive to established qualitative differences in seizure severity. Our results also suggest that seizure severity is modulated over different timescales. We envisage that our proposed seizure severity library will be expanded and updated in collaboration with the epilepsy research community to include more measures and modalities.

\subsection*{Key Points} 
\begin{itemize}
    \item Existing measures of seizure severity can be complemented by objective quantitative markers of seizure EEG severity. 
    \item EEG-based markers of seizure severity can distinguish clinically distinct seizure types. 
    \item Quantitative severity markers can be used to investigate fluctuations in seizure severity over time in individual patients. 
\end{itemize}

\newpage
\section{Introduction}
Seizure severity is an important clinical measure for patients with epilepsy that is strongly correlated with quality of life \citep{bautista2009seizure}. However, the best approach for measuring seizure severity remains unclear. Existing scales for measuring seizure severity, including the National Hospital Seizure Severity Scale (NHS3) \citep{duncan1991chalfont,o1996national}, the Liverpool Seizure Severity Scale (LSSS) \citep{baker1991development}, and the Seizure Severity Questionnaire (SSQ) \citep{cramer2002development}, are composed of questions on various aspects of seizures including warnings, ictal and postictal phenomena, and resultant injuries. Most scales separate seizures by their clinical classification \citep{cramer2001quantitative} to reflect differences in severity across different seizure types. 

A primary shortcoming of existing measures of seizure severity is their reliance on patient or carer recollection \citep{cramer2001quantitative}. For example, a patient's recollection of their seizure may be impaired as a result of the seizure itself \citep{dubois2010seizure, tatum2001outpatient}. It is hence challenging to assess changes in severity from seizure-to-seizure in an unbiased manner for the full range of a patient's seizures. Objective, quantitative tools for measuring severity of individual seizures are therefore needed to understand variations in seizures on different timescales.

EEG-based severity markers are a potential approach to quantifying seizure severity. Past studies have used EEG features such as ictal duration \citep{ochoa2021seizure} and spatial synchronisation \citep{ravan2016quantitative} as proxies for seizure severity. The anatomical spread of seizure activity has also been suggested as a measure of seizure severity \citep{cramer2001quantitative}. It is yet to be determined how such measures compare and which to use for each individual patient. 

Moreover, various seizure features, which are directly associated with severity, fluctuate over time. For example, focal seizures are more likely to generalise in sleep \citep{Jobst2001}, particularly in temporal lobe epilepsy \citep{bazil1997effects}. The extent of postictal suppression also depends on the time of day of seizure occurrence \citep{Lamberts2013, Peng2017}. Subclinical seizures (without clinical symptoms) also follow circadian patterns \citep{jin2017prevalence}. Furthermore, it has been shown that `seizure spatiotemporal evolutions'  \citep{schroeder2020seizure} and ictal onset dynamics \citep{saggio2020} differ within individuals on circadian or longer timescales. Therefore, monitoring fluctuations in seizure severity could lead to a better understanding of an individual's epilepsy. 

To objectively quantify seizure severity we provide an expandable library of interpretable EEG-based markers of seizure severity. As a way of validation we test if seizure severity markers distinguish clinically distinct seizure types \citep{fisher2017instruction} with known differences in severity. We further show that markers of seizure severity are patient-specific. As a proof of principle we further demonstrate fluctuations in severity over circadian or longer timescales. 

\newpage

\section{Methods}
\subsection{Patient selection and data acquisition}
This retrospective study analysed iEEG recordings of 1009 seizures across 63 patients undergoing pre-surgical evaluation for medically refractory epilepsy. Seizure types were labelled by clinical teams according to ILAE classifications: 656 focal, of which 232 were focal aware, 176 were focal impaired awareness, 248 were uncategorised; 323 subclinical; 6 focal to bilateral tonic-clonic (FTBTC). Within this work, seizures with focal onset, clear clinical correlates, and no propagation to the contralateral hemisphere were labelled as focal seizures. Suppl. Section~\ref{suppl:metadata} provides more details on the patient cohort. 

Data were collected from two epilepsy monitoring units (EMUs) in the UK: UCLH and Glasgow with 49 and 14 patients, respectively. Anonymised iEEG recordings were analysed following approval of the Newcastle University Ethics Committee (reference number 17042/2021). Electrographic seizure start and termination were labelled by clinical teams. Ictal periods were extracted with two minutes of pre- and post-ictal activity. 

\subsection{iEEG Pre-processing}
We first downsampled all EEG to 256Hz. Pre-ictal noise was detected using an iterative noise detection algorithm and visual inspection; noisy channels were removed from all seizures (see Suppl. Methods~\ref{noisedetect}). The iEEG was re-referenced to a common average reference, notched filtered at 50 and 100Hz (2Hz window) to remove line noise, and band-pass filtered between 0.5 and 100Hz (fourth order, zero phase shift Butterworth). 

\subsection{Seizure Markers}
The selection of markers was inspired by seizure detection literature (e.g., \cite{alotaiby2014eeg, guo2010automatic, birjandtalab2016nonlinear}). To quantify different types of features our library of objective seizure severity markers has three main branches: 
\begin{itemize}
    \item `peak' markers to measure the peak level of activity that occurs during a seizure 
    \item `spatial' markers to summarise spread of ictal activity across recording channels
    \item `suppression' markers to evaluate post-ictal suppression 
\end{itemize}
Ictal duration was also included as an additional severity marker \citep{beniczky2020biomarkers}. Supplementary Table ~\ref{tab:severity_markers} gives detailed mathematical definitions of all markers.

Common notation is used throughout the definition of markers: $x$ is the time series for one channel, $k$ is the time point, $N$ is the number of time points in the segment, $C$ is the number of recording channels, and $T$ is the seizure duration in seconds.

\subsubsection{Peak Markers}
\begin{figure}
    \centering
    \includegraphics{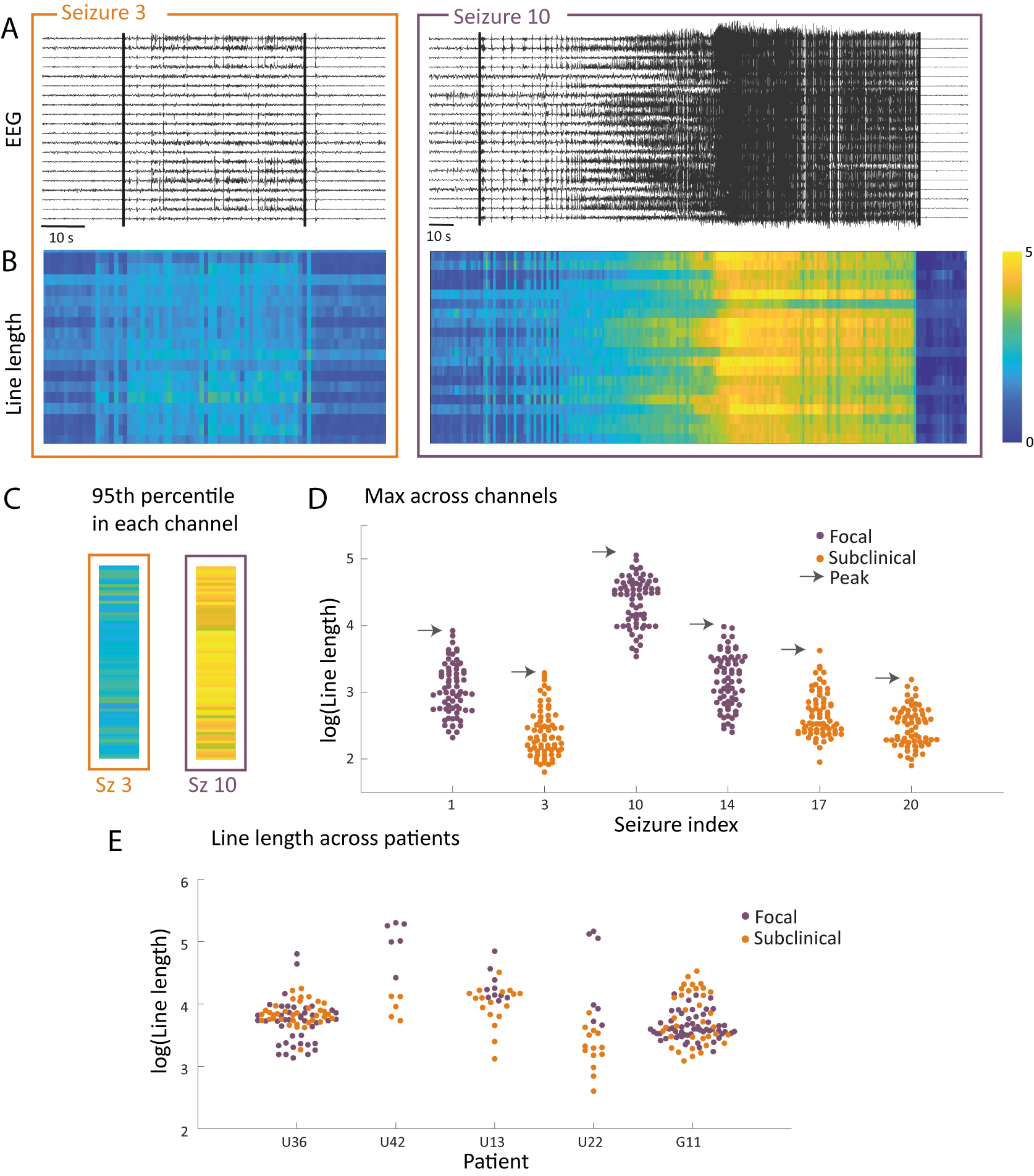}
    \caption{\textbf{Visualising the workflow for calculating peak markers for example patient U22.}\\ A) Intracranial EEG traces for a subclinical (orange) and focal (purple) seizure in an example patient, with a subsection of recording channels for visualisation. B) Heat-maps of the line length marker in one second epochs for seizures in A. C) $95^{th}$ percentile of line length measures for each channel across time. D) Bee-swarm representation of the same data as C, also for a few more example seizures in this patient. Grey arrows point to the maximum value across channels, this is the peak value for the seizure. E) Log-transformed peak line length values (maximum channel value across $95^{th}$ percentiles), as indicated by grey arrows in D in five example patients, each data point represents a seizure.}
    \label{fig:figure_1}
\end{figure}

The maximum level of activity in the ictal phase was estimated using peak markers of the iEEG features: line length \citep{olsen1994automatic}, energy \citep{hamad2016feature}, and band-power \citep{acharya2013automated} (in $\delta$ (1-4Hz), $\theta$ (4-8Hz), $\alpha$ (8-13Hz), $\beta$ (13-30Hz), low-$\gamma$ (30-60Hz) and high-$\gamma$  (60-100Hz) bands), each of which have previously been used within seizure detection algorithms \citep{birjandtalab2016nonlinear, boonyakitanont2020review}. Each seizure recording (Fig.~\ref{fig:figure_1}A) was separated into one-second epochs with no overlap from which each peak marker was calculated; resulting in eight $T\times C$ matrices (Fig.~\ref{fig:figure_1}B).

For each severity marker (i.e., each matrix) we first summarised markers across time; for each recording channel, the $95^{th}$ percentile of each marker was calculated (Fig.~\ref{fig:figure_1}C). The maximum value across channels was then used as the estimated peak activity of the seizure (Fig.~\ref{fig:figure_1}D). As expected, markers differ across seizure types and patients (Fig.~\ref{fig:figure_1}E). Once summarised over time (by the $95^{th}$ percentile of each channel across time) and across channels (maximum value), we log-transformed the measures to normalise their distributions.

\subsubsection{Spatial Markers}
\begin{figure}
    \centering
    \includegraphics[width=1\textwidth]{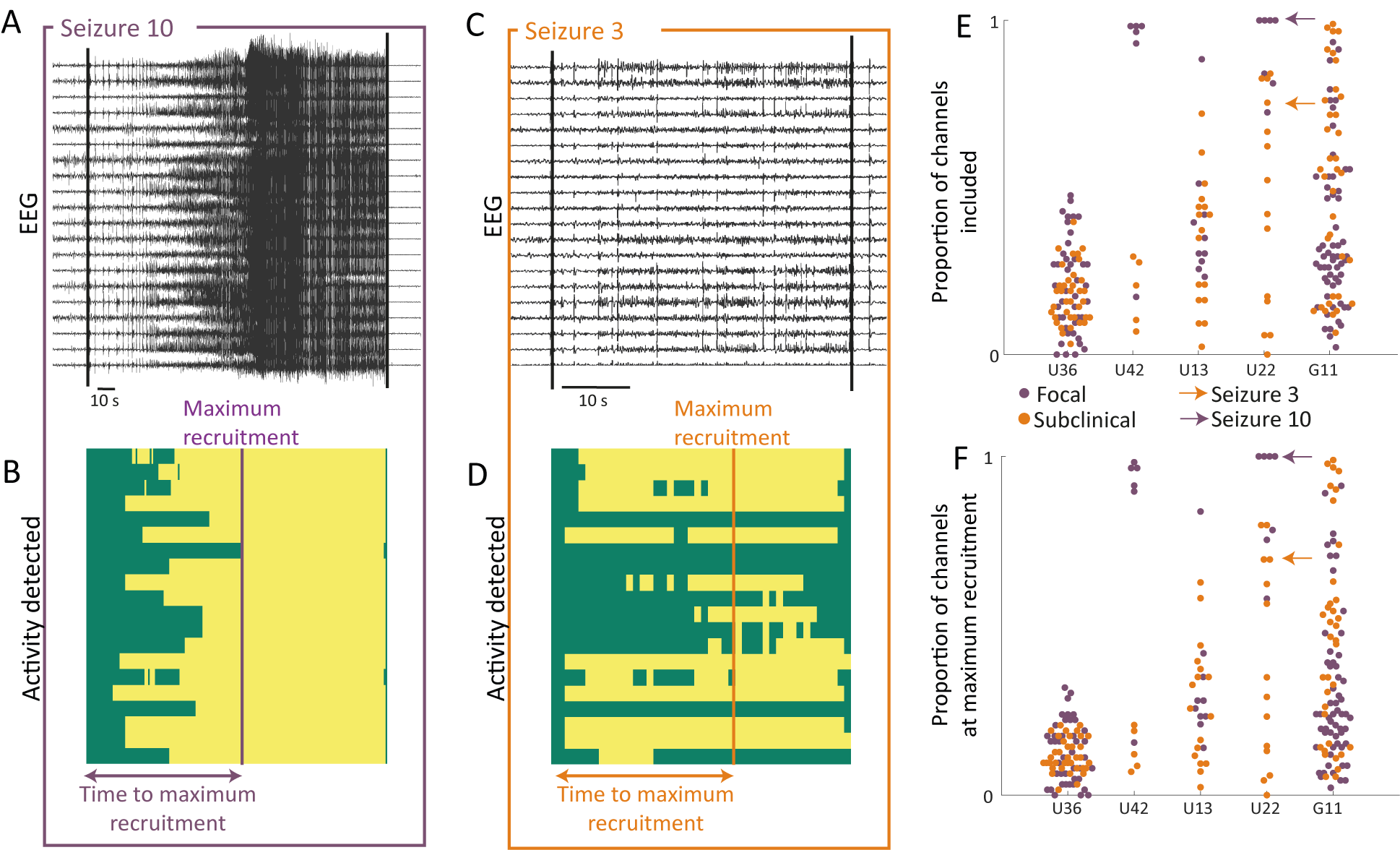}
    \caption{\textbf{Visualising spatial markers for example patient U22.}\\ A \& C) Intracranial EEG traces of an example focal/subclinical seizure with a subset of recording channels. B) Corresponding binary map of seizure imprint (yellow indicates seizure activity, green no seizure activity) across time in the same subset of channels as in A \& C. E) Swarm plot of the proportion of channels with seizure activity at any point in the seizure for all seizures in five example patients. F) Swarm plot of the proportion of channels with seizure activity at the point of maximum recruitment for all seizures for five example patients.}
    \label{fig:figure_2}
\end{figure}

The extent of the spread of ictal activity across recording channels was captured through spatial markers. For each channel, baseline (pre-ictal) and ictal recordings were divided into 1 second, non-overlapping windows, from which each of eight features (line length, energy, 6$\times$ band-powers) were calculated. Seizure activity was algorithmically detected based on abnormality (median absolute deviation (MAD) scores) relative to the pre-ictal period in each of the eight feature matrices. For each window per channel an MAD score greater than five in any of the eight features suggested potential seizure activity. An additional step (see Suppl \ref{spatialmarkers} for details) prevented spurious non-seizure activity being detected (e.g., caused by noise or a short spike). This algorithm yielded a binary map identifying channels and time windows with seizure activity during the ictal period. We term this matrix the `imprint' of the seizure (see Fig~\ref{fig:figure_2}A and C for EEGs and B and D for corresponding imprints).

Four markers were derived from the seizure imprint: the proportion of channels with seizure activity at any point in the ictal phase (Fig.~\ref{fig:figure_2}C, example patients), the proportion of channels with simultaneous seizure activity at the point of maximum recruitment (Fig.~\ref{fig:figure_2}D, example patients), the time taken from seizure onset to the time of maximum recruitment, and the proportion of the seizure duration taken to reach maximum recruitment.

\subsubsection{Suppression Markers}
\begin{figure}
    \centering
    \includegraphics[width=\textwidth]{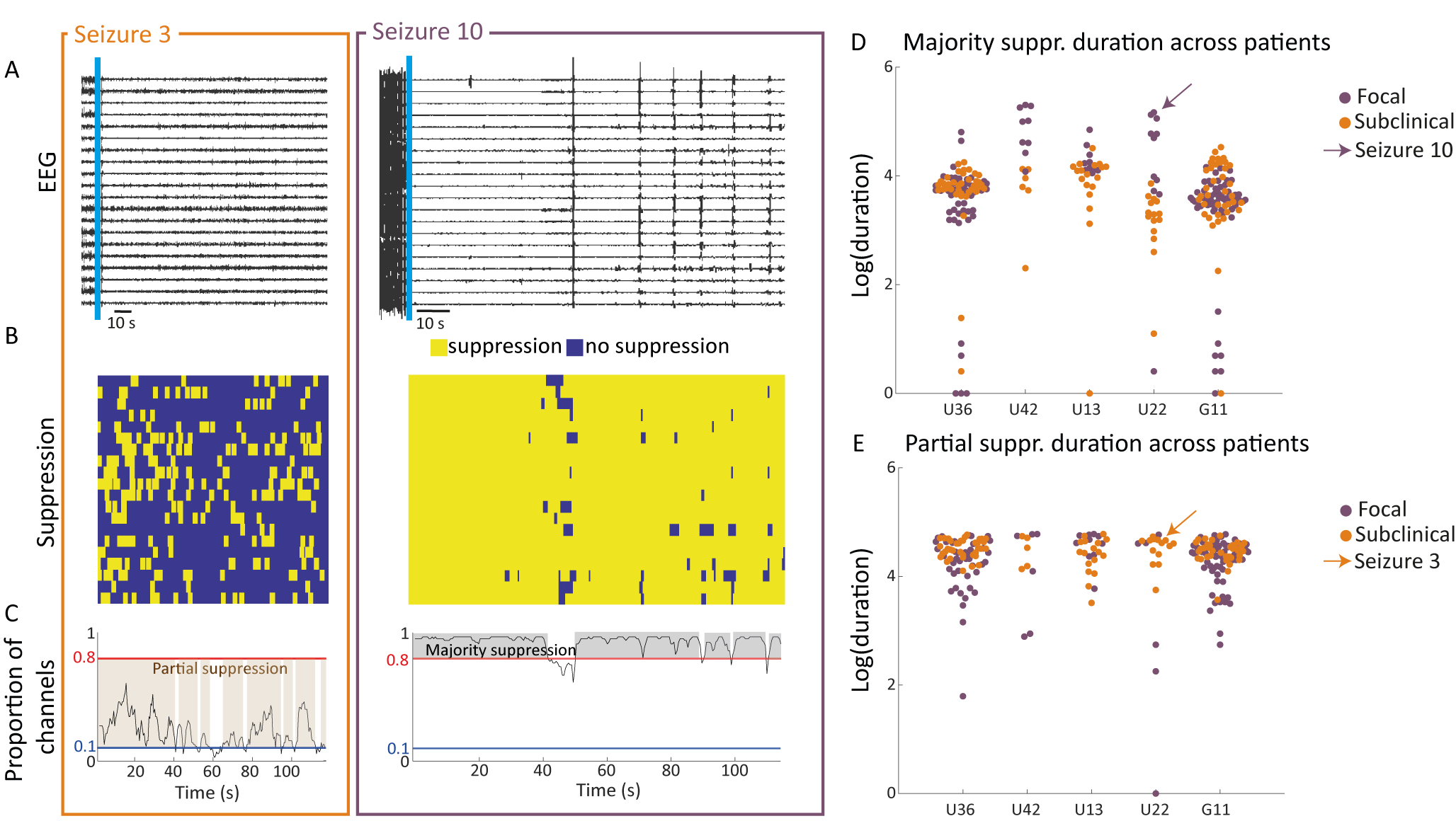}
    \caption{\textbf{Visualising suppression markers for example patient U22.}\\
     A) Intracranial EEG traces of example subclinical (orange) and focal (purple) post-ictal segments with a subset of recording channels. B) Corresponding binary maps of channels with suppression ($<5\%$ of preictal activity levels) in the same subset of recording channels. C) Proportion of suppressed channels across 120 seconds of postictal activity. Segments of majority suppression and partial suppression are highlighted. D) Swarm plot of (log-transformed) majority suppression duration for all seizures for five example patients. E) Swarm plot of (log-transformed) partial suppression duration for all seizures for five example patients.}
    \label{fig:figure_3}
\end{figure}

Duration and strength of post-ictal suppression was captured by our suppression markers. Signal range was computed in 0.5-second non-overlapping windows. For each channel post-ictal ranges were compared against the distribution of preictal ranges. Ranges below the fifth percentile of preictal range were labelled as suppressed (see Fig.~\ref{fig:figure_3}A for a post-ictal EEG and its corresponding suppression matrix in B). Periods of suppression were labelled as majority suppression or partial suppression based on the proportion of suppressed channels (Fig.~\ref{fig:figure_3}C). Durations of majority suppression and partial suppression (Fig.~\ref{fig:figure_3}D \& E) were calculated using a 2.5-second moving sum to account for short spikes of activity in suppressed segments. Further details are provided in Suppl. Methods~\ref{suppressionmarkers}. 
The suppression duration was computed as the time following seizure offset with a one-second buffer. A third suppression marker, suppression strength, was defined as the median proportion of channels with suppression across the duration of the post-ictal recording. 
Whilst we analysed 120 seconds of postictal activity, duration of suppression may have exceeded this 120s \citep{ochoa2021seizure}. Therefore, suppression durations of 120s in the following should be understood as `at least 120 seconds'. 

\subsection{Statistical analysis}
Statistical analyses were then performed in RStudio. P-values were calculated for reference and visualisation, not to stratify patients for further analyses. 

\subsubsection{Validating markers against ILAE seizure classification} 
International League Against Epilepsy (ILAE) seizure classification \citep{fisher2017operational} was used as a validation for seizure severity. Our main analyses compared focal \textit{vs.} subclinical seizures and focal aware \textit{vs.} impaired awareness seizures; supplementary analyses are shown comparing focal \textit{vs.} FTBTC seizures. Performance of markers was assessed through how well they distinguish these seizure types. We applied two strategies for validation, across and within patients, to separately assess performance of markers in distinguishing clinically distinct seizure types. 

\paragraph{Across Patients}
For each marker three hierarchical logistic regression models were compared to assess marker and/or patient effects. Specifically, we created a model considering only random patient effects and considering both fixed marker effects and random patient effects (random intercept \& random intercept and slope models). The fit of each model was assessed using Akaike information criterion, Bayesian information criterion, and deviance. Models with poor fit were deemed inadequate and removed. Assumptions of logistic regression models were checked for each model individually. The quality of each model as a classifier of seizure type was assessed through the area under the curve (AUC) for receiver operating characteristic (ROC) curves with 100 decision thresholds. Performance was assessed based on AUC thresholds (AUC$>0.7$ is acceptable, $>0.8$ is excellent, and $>0.9$ is outstanding) \citep{mandrekar2010receiver}. Supplementary analyses are shown for focal \textit{vs.} FTBTC seizures (Suppl. Table~\ref{tab:across_focal_ftbtc}) and focal \textit{vs.} subclinical seizures in TLE (Suppl. Table ~\ref{tab:tle_across_focal_subclin}) and eTLE (Suppl. Table ~\ref{tab:etle_across_focal_subclin}).

\paragraph{Within Patients}
Each marker's performance in distinguishing seizure types for each patient was assessed using two-tailed Wilcoxon rank sum tests. Patients were included in within-patient validation if that they had a minimum of five seizures, with two or more seizures of each type. The distinction between markers of different seizure types was quantified using the effect size ($r$) calculated as:
$$r = \frac{Z}{\sqrt{N}}$$
where $Z$ is the $Z$-statistic and $N$ is the total sample size.  The $r$ value was bounded between zero and one with values closer to one indicating larger effects. It is common in the literature to consider $0.1\leq r<0.3$ as a small effect, $0.3\leq r<0.5$  as a moderate effect, and $r\geq$ 0.5 as a large effect. 

\subsubsection{Circadian and longer-term modulation of seizure severity}
We additionally assessed circadian and longer-term fluctuations in seizure severity. For individual patients we assessed circadian fluctuations using rank circular-linear correlation \citep{mardia2000directional} using the \textit{cylcop} R package \citep{cylcop}. P-values were calculated through a permutation test with 1000 permutations. Inclusion criteria were that patients must have 20 or more recorded seizures irrespective of the frequencies of each seizure type. This threshold was chosen based on performance of circular-linear correlation on simulated data with varied sample sizes and noise. Long-term fluctuations in severity were assessed using Spearman's rank correlation between markers and the time since first recorded seizure. 

\subsection{Code and data availability}

The analysis code and data are available on Zenodo.org upon acceptance. The expandable library of severity markers is already available on GitHub(\url{https://github.com/cnnp-lab/seizure_severity_library}), and we invite contributions from the community.

\section{Results}
We computed each of the 16 proposed seizure severity EEG markers for all 1009 recorded seizures. We first validated each marker by assessing performance in distinguishing different ILAE classification both across all patient seizures and within each patient. However, we envisage additional uses of this library and, as an example, demonstrate its potential ability to detect fluctuations in seizure severity over time. 

\subsection{Severity markers distinguish between ILAE clinical seizure types across patients and seizures}
\begin{figure}
    \centering
    \includegraphics[width=1\textwidth]{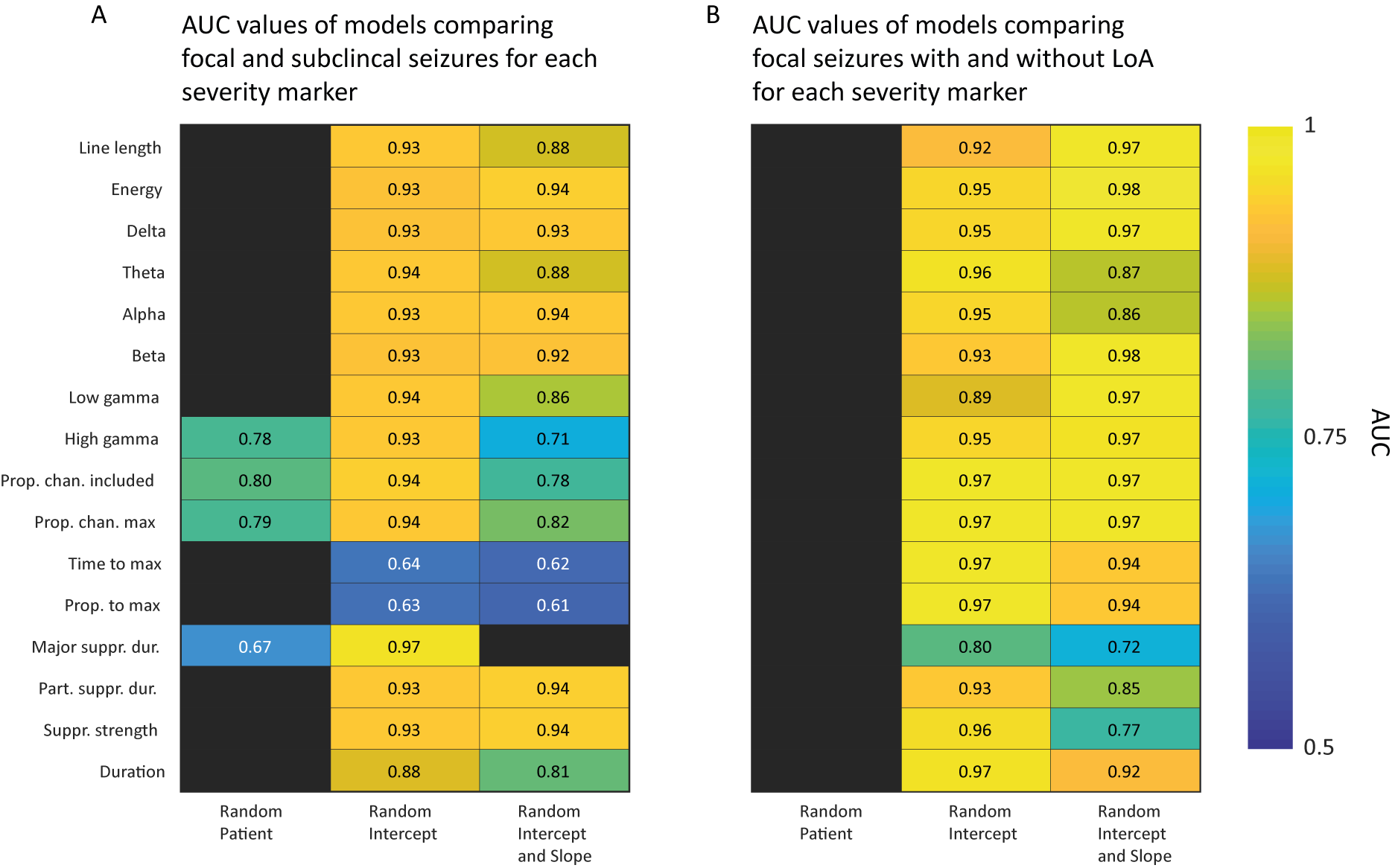}
    \caption{\textbf{Validating markers against ILAE classification across patients.}\\ 
   A) Heat-map of AUC values for hierarchical logistic regression models comparing focal and subclinical seizures. B) Heat-map of AUC values for hierarchical logistic regression models comparing focal seizures with and without loss of awareness (LoA).
   }
    \label{fig:figure_0}
\end{figure}

To validate our markers we assessed their ability to distinguish focal \textit{vs.} subclinical seizures and focal seizures with and without impaired awareness across patients. Specifically, for each of the 16 markers, we compared seizure types across all patients using hierarchical mixed effects logistic regression models. Fig.~\ref{fig:figure_0} displays the AUC values obtained for each model in all markers when comparing focal \textit{vs.} subclinical (A) and focal seizures with and without impaired awareness (B). There were clear patient differences in the marker values; however, the majority of models created with only patient effects were unacceptable classifiers (AUC $<$ 0.7 or model assumptions not met), suggesting that between-patient differences alone did not account for differences between focal and subclinical seizures. In contrast, 14 severity markers yielded excellent classifier performance with random intercept models or random intercept and slope models. As seizure duration is often used to assess seizure severity \citep{beniczky2020biomarkers} we compared the performance of each marker against the performance of duration in distinguishing seizure types (see Fig.~\ref{fig:suppl_auc} C$\&$ D) using a bootstrapping procedure (see Suppl. methods \ref{bootstrap}). When comparing focal \textit{vs.} subclinical seizures observed AUC values for all markers (except time and proportion of seizure to maximum recruitment) were larger than most of the distribution of AUC values for seizure duration. When comparing focal seizures with and without impaired awareness all peak markers except theta and alpha band-powers, and all spatial markers outperformed seizure duration.\\
Supplementary material ~\ref{acrosspat} shows additional results comparing focal \textit{vs.} FTBTC seizures and comparing focal \textit{vs.} subclinical seizures in TLE and eTLE. When comparing focal \textit{vs.} FTBTC seizures, all markers created excellent or outstanding classifiers through random intercept models. We further subdivided patients into those with temporal lobe epilepsy (TLE) and extra temporal lobe epilepsy (eTLE) from which we repeated across patient analyses for focal \textit{vs.} subclinical seizures. When separating patients into TLE and eTLE only sample sizes were 360 and 595 seizures, respectively. For TLE patients five markers showed excellent or outstanding performance in random intercept models (Suppl. Table~\ref{tab:tle_across_focal_subclin}). For eTLE patients all markers had excellent or outstanding performance in random intercept and/or random intercept and slope models (Suppl. Table~\ref{tab:etle_across_focal_subclin}).
\subsection{Severity markers distinguish between ILAE clinical seizure types within patients}
We next validated our markers by quantifying distinctions between ILAE seizure types within individual patients. We analysed effect sizes between seizure types using Wilcoxon rank sum test $r$-values. Using our inclusion criteria, we could compare focal and subclinical seizures in 15 patients. Patients included in this analysis did not differ in demographics (sex, age, disease duration, and epilepsy diagnosis) relative to the entire cohort. Majority suppression duration could not be validated as many patients did not have sufficient seizures with periods of majority suppression.

\begin{figure}
    \centering
    \includegraphics[width=1\textwidth]{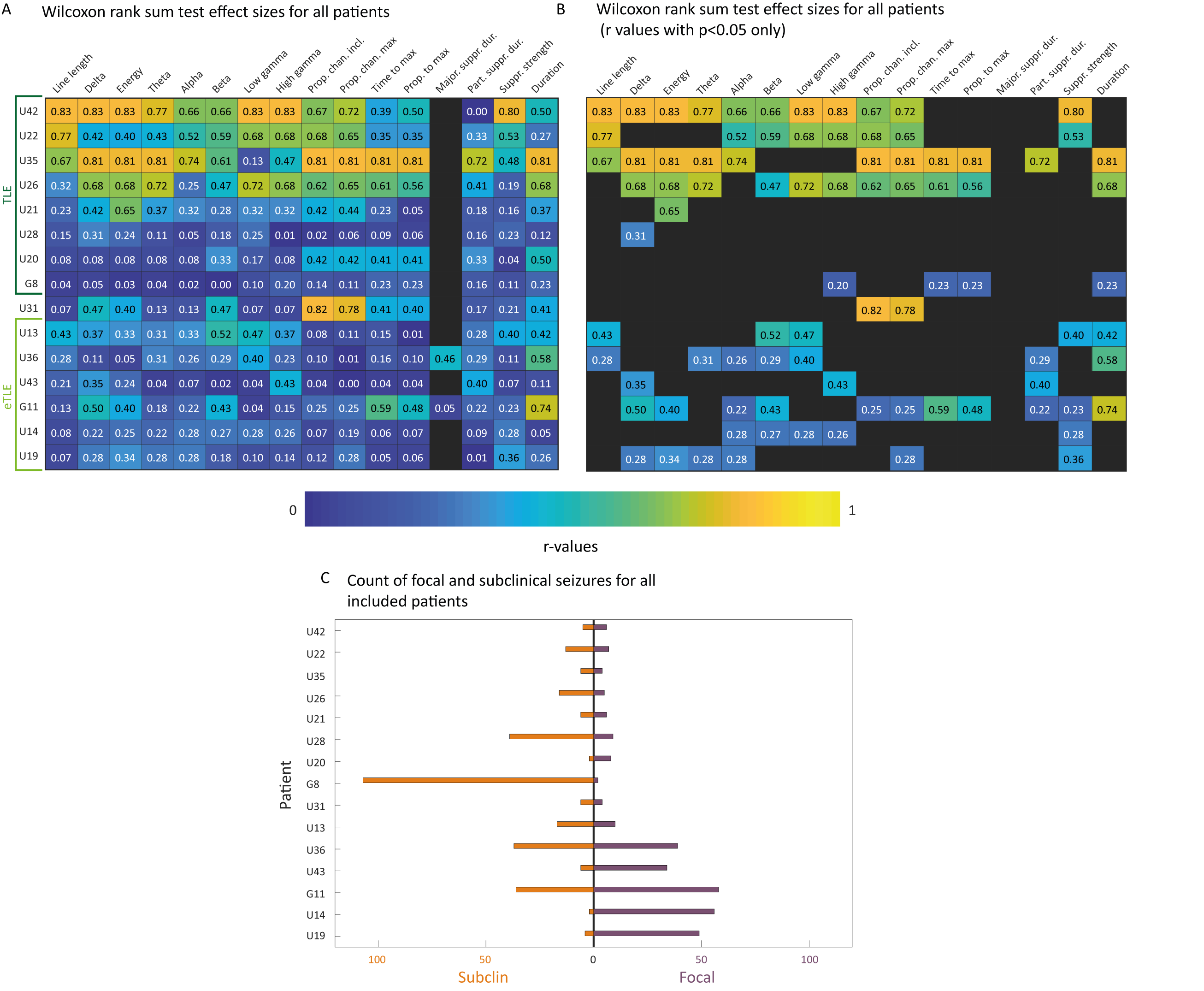}
    \caption{\textbf{Validating markers against ILAE classification on a within-patient basis.}\\
    A) Wilcoxon rank sum test $r$-values obtained through comparing focal and subclinical seizures. Each row is a patient, and each column is a marker. Patients were sorted by descending $r$-values within the TLE and eTLE groups. B) Same as A, filtered by $p<0.05$. C) Paired bar chart displaying counts of focal and subclinical seizures for each patient included in within-patient validation.
}
    \label{fig:figure_4_B}
\end{figure}

Moderate to large effects ($r>0.3$, $p<0.05$) in three or more markers were found for eight of the 15 included patients (53.3$\%$). The heat-maps of $r$-values is shown in Fig.~\ref{fig:figure_4_B}A, Fig.~\ref{fig:figure_4_B}B shows a heat-map of $r$-values only where $p<0.05$. The number of focal and subclinical seizures recorded per patient varied (see Fig.~\ref{fig:figure_4_B}C). Effects were notably higher in four patients, all of whom were TLE patients, supporting that performance of markers is likely patient-specific. We investigated the effect of various other patient metadata (sex, TLE/eTLE, surgical outcome, disease duration, age, number of recording channels, and number of recorded seizures) on marker performance in Suppl. Table ~\ref{tab:com_eff_size}. Most notably, there was a large effect between spatial markers for patients with TLE compared to eTLE, but none of the other patient features showed consistent or noteworthy effects. 
Comparing performance of our markers against seizure duration, in five patients (33$\%$), duration alone was not a useful marker of seizure severity ($r<0.3$, $p>0.05$). However, in each of these patients, at least three other markers were useful ($r>0.3$, $p<0.05$) in distinguishing focal and subclinical seizures

\subsection{Seizure severity changes across different time scales}
\begin{figure}
    \centering
    \includegraphics[width=0.75\textwidth]{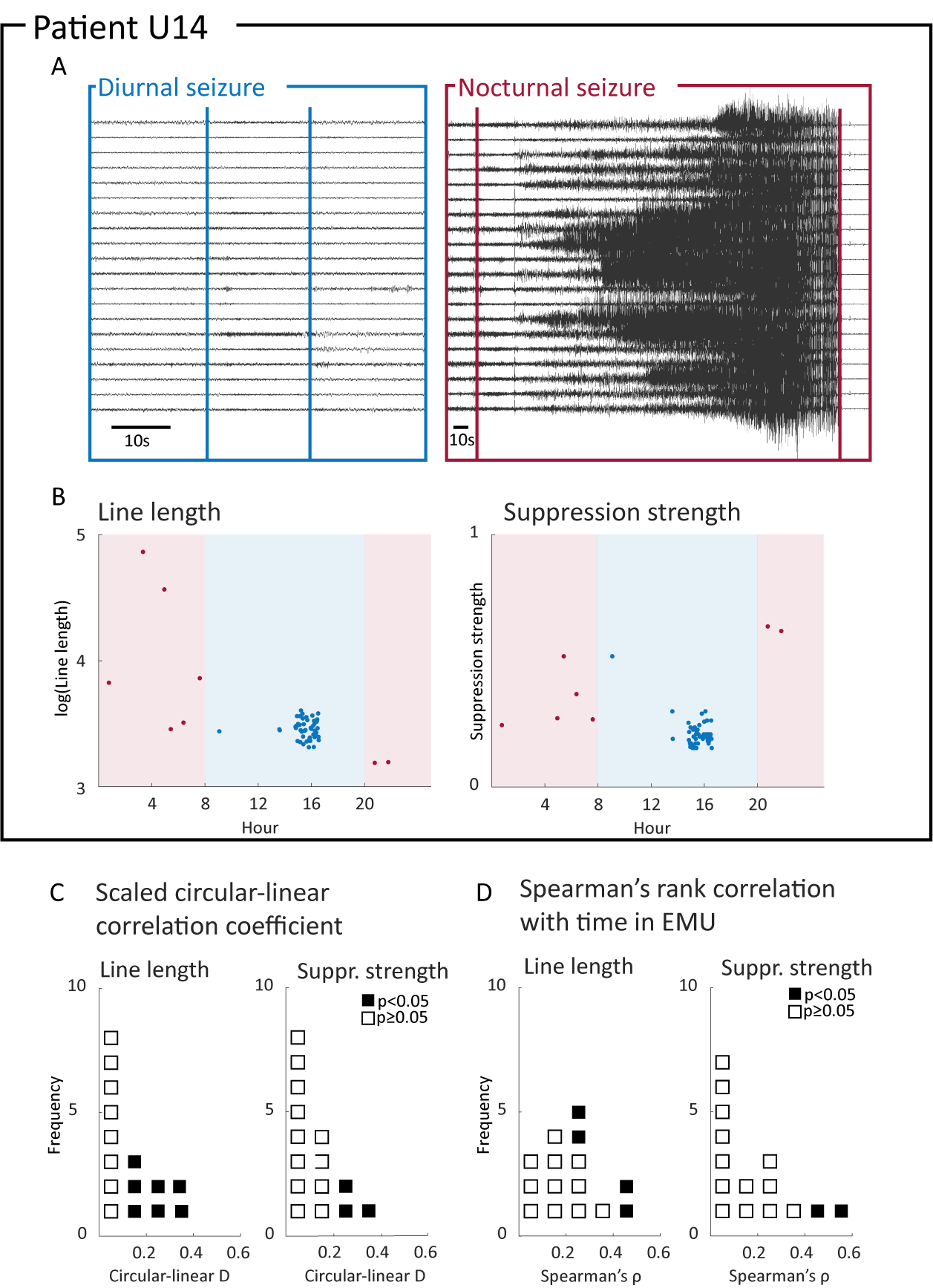}
    \caption{\textbf{Detecting circadian and longer-term modulation of seizure severity.}\\
    A) iEEG recordings for a day-time (blue) and night-time (pink) seizure from example patient U14. B) Plot of marker against time of day for line length and postictal suppression strength. Pink background indicates evening/night, whilst blue background indicates daytime.  C) Dot plot of scaled circular-linear correlation coefficients between markers and time of day across included patients. P-values $<$ 0.05 obtained through permutation tests are highlighted in black. D) Dot plot of absolute Spearman's rank correlation coefficient between markers and time in EMU across included patients. Correlations with p-values $<$ 0.05 are highlighted in black.}
    \label{fig:figure_5}
\end{figure}

Finally, we used our markers to capture fluctuations in seizure severity on circadian and longer timescales in 15 patients. Fig.~\ref{fig:figure_5}A shows example day-time and night-time seizure iEEG traces from the same patient, U14. In U14, seizures occurring at different times of day appeared to have different characteristics; for example, line length and suppression strength differences are higher in nocturnal seizures (Fig.~\ref{fig:figure_5}B and C). The association between these markers and seizure times, was measured using circular-linear correlation \citep{mardia2000directional}. Eight patients (66.7$\%$) had correlations with $\rho > 0.2$ and $p < 0.05$ for at least three markers. 

We additionally asked if our severity markers also changed over the span of each patient’s recording. Fig.~\ref{fig:figure_5}D shows the absolute Spearman’s rank correlation between two example markers and the time of each seizure relative to the start of the recording. This measure captures the strength, but the not direction, between marker values and the time of seizure occurrence. In eight out of 15 patients (53.3$\%$), at least three markers had correlations with $\rho > 0.3$ and $p < 0.05$ with the amount of time elapsed since the start of the recording. 

Correlation coefficients for all markers are shown in Supplementary Tables  ~\ref{tab:circ_lin_peak} - ~\ref{tab:cor_emu_other}. Moderate to strong correlations can be seen in a wide range of markers and patients; thus, we conclude that circadian and longer-term changes in EEG severity can be detected in the majority of patients.\\
We were limited by the time spent in the EMU; therefore, our findings on modulation are proof-of-concept. These results should be interpreted as evidence that our markers could be used to capture fluctuations in severity. 

\section{Discussion}
We evaluated 16 objective quantitative markers of seizure severity derived from iEEG recordings of patients with refractory focal epilepsy. Our goal was to offer a collection of markers which can be used as output measures for clinical trials, tracking fluctuations in seizure severity, or other applications. Our results demonstrated that almost all severity markers could distinguish focal \textit{vs.} subclinical seizures across our cohort of 63 patients. Importantly, marker performance was patient-specific, indicating that different groups of patients are best evaluated with a subset of our proposed markers; thus, our approach of providing a severity library for future work to draw from is an important contribution. We also found that severity fluctuated on circadian and longer-term timescales in a patient-specific manner supporting the use of EEG-based severity markers to investigate temporal modulation of seizure severity. Our work may therefore also facilitate personalised, time-adaptive treatments or enhance our understanding of the chronobiology of seizures. 

Existing scales of seizure severity have been used as outcome measures in clinical trials \citep{beenen1999comparative, dagar2011epilepsy, kverneland2018effect, szaflarski2018cannabidiol}. However, scales depend on patients’ ability to recall seizures over weeks \citep{baker1991development, cramer2001quantitative} leading to concern over their reliability. Many scales also focus on patient risk rather than objective severity. For example, the NHS3 stipulates that seizures occurring in bed are automatically scored zero for falls, potentially underestimating their electrographic and neurobiological severity. No existing scales assess individual seizure severity in an objective quantitative manner, making small changes in severity difficult to capture. Our library of quantitative EEG markers addresses these limitations, providing a complementary approach for measuring and understanding seizure severity. 

Our approach of validating our markers was to compare two seizure types that have obvious distinctions in terms of their neurobiological and symptomatic severity: namely subclinical \textit{vs.} focal seizures. The proportion of subclinical \textit{vs.} focal seizures within this data (323 \textit{vs.} 656) agrees with previous literature \citep{farooque2014subclinical}, suggesting that our seizure type labels are not biased. 
Previous literature suggests that subclinical and focal seizures have different EEG features \citep{blume1984eeg}, even within the same patient \citep{farooque2014subclinical}, thus making it a good standard to compare to. However, our proof-of-principle validation against seizure type is only one of many possible standards; future work could test other standards that are tailored to the research question.


One main finding of this work was that the performance of seizure severity markers derived from iEEG recordings is highly patient-specific. Peak markers tended to perform well as did some spatial markers (proportions of channels measures). The remaining markers varied in their performance, even among patients with better distinctions based on other markers. Results suggest that spatial markers have the highest performance in distinguishing focal seizures with and without impaired awareness. We suggest testing the entire library of markers for each new patient to determine which, if any, are the most appropriate for the desired application.  

Different aspects of seizure severity have been repeatedly reported to follow circadian, sleep/wake, and longer timescale modulations. For example, secondary generalisation and post-ictal suppression occur more often in seizures arising from sleep \citep{Jobst2001,Lamberts2013,Peng2017}. Subclinical seizures are also reported to follow a circadian pattern \citep{jin2017prevalence}. Recent studies also reported modulations at circadian and longer timescales within many patients in terms of seizures electrographic evolutions \citep{schroeder2019slow,panagiotopoulou2020fluctuations} and other seizure properties \citep{schroeder2022chronic}. In agreement with previous literature, we found evidence that EEG based seizure severity markers are modulated on circadian and longer timescales although the effect size of the modulation is patient-specific and weak in some patients. We suggest that, similar to previous work \citep{panagiotopoulou2020fluctuations}, capturing data of the potential modulations and directly relating those to the severity markers in a multivariate model may be insightful.


Limitations and future work: The patients included in this study are presurgical candidates with refractory focal epilepsy; therefore, our library needs to be expanded and tested in other epilepsy syndromes. The use of iEEG allows for good signal quality but does not capture activity beyond a small part of the brain. Electrode placement was determined by clinical need and therefore the location of electrodes varied across patients. This variability means that spatial markers do not represent the same information in different patients and thus hierarchical statistical approaches are needed to compare markers across patients. Future work could use simultaneous scalp and intracranial EEG to validate markers of spread based on iEEG in different anatomical regions. Within this work spatial markers based on activity in regions of interest (ROIs) rather than individual channels was considered, unfortunately electrode location was not available for all patients. We opted to maintain our channel-based spatial and suppression markers to maintain our sample size. Further research including a larger cohort with available electrode location information is required to confirm if spatial markers derived from ROIs could be used to capture seizure severity. Our methods could be extended to sub-scalp EEG with some alterations to account for lower spatial coverage. Although the lower coverage presents a challenge, previous studies suggest encouraging findings. For example, \citep{parvez2015epileptic} predicted seizure occurrence using only six recording channels per patient. Furthermore, recordings from only 16 locations on the surface of the brain captured critical slowing \citep{maturana2020critical} giving evidence that alterations in EEG around seizures can be captured with few electrodes. Extension of our library to scalp EEG and other modalities are planned and with our open code-base on GitHub we welcome contributions from the community.\\
As recordings took place in EMUs, patients were also under non-normal conditions during recordings; anti-seizure medications (ASMs) are often tapered, and patients are potentially under an increased amount of stress. Future work might use continuous recordings to capture the full range of interictal brain dynamics to better estimate spatial and suppression properties of seizures. Future work should also investigate the three-way relationship between severity markers, seizure type, and circadian influences. Further, electrographic activity can fluctuate for weeks following electrode implantation \citep{ung2017intracranial}; although, the pre-ictal baseline that we applied for spatial and suppression markers may render those markers less sensitive to such fluctuations. Future work needs to disentangle the biological, technological, and pathological influences on EEG biomarkers; this remains an open challenge for various applications.
the results of this work may have been influenced by such fluctuations, especially in modulation analyses. Regardless, our results remain meaningful as a proof-of-concept that our markers can be used to detect fluctuations in ictal electrographic activity and, by extension, seizure severity. 


\section{Conclusion}
In conclusion, we propose 16 EEG markers of seizure severity which can be used to complement existing measures. Most markers were validated against ILAE classification on an across patients basis. Marker performance, as measured by their ability to distinguish seizure types and capture fluctuations in seizure severity, is strongly patient-specific. We also detected circadian and longer timescale fluctuations in seizure severity which may be relevant for a range of applications including capturing treatment response and seizure forecasting  \citep{takahashi2012state, cook2016human, freestone2017forward}. Our library therefore contributes to ongoing efforts in characterising seizures over time, seizure prediction, and generally designing novel, personalised treatment plans that manage and mitigate severe seizure.


\section*{Author Contributions}
\begin{itemize}
  \item Conceptualization: SJG LW GMS MP PNT YW
  \item Methodology: SJG LW GMS MP CP YW

\item Software/validation: LW YW

\item Formal analysis: SJG MP YW

\item Resources: FC AC BD JSD JF VL SL 

\item Data curation: JB RF GMS YW

\item Writing: SJG LW GMS MP JSD CP KW PNT YW

\item Supervision: YG RHT KW PNT YW
\end{itemize}

\section*{Acknowledgements}
We thank members of the Computational Neurology, Neuroscience \& Psychiatry Lab (www.cnnp-lab.com) for discussions on the analysis and manuscript. S.J.G and M.P. are supported by the Engineering and Physical Sciences Research Council (EP/L015358/1) and ADLINK; P.N.T. and Y.W. are both supported by UKRI Future Leaders Fellowships (MR/T04294X/1, MR/V026569/1). J.S.D. is supported by the Wellcome Trust Innovation grant 218380. J.S.D., J.T. are supported by the NIHR UCLH/UCL Biomedical Research Centre. None of the authors has any conflicts of interest to disclose.

\end{doublespace}
\newpage

\bibliography{refs}

\newpage


\renewcommand{\thefigure}{S\arabic{figure}}
\renewcommand{\thetable}{S\arabic{table}} 
\counterwithin{figure}{subsection}
\counterwithin{table}{subsection}
\renewcommand\thesubsection{S\arabic{subsection}}
\setcounter{subsection}{0}

\section*{Supplementary}
\subsection{Glossary of acronyms}
\textbf{Clinical terms}\\
\textbf{ASMs:} Anti-seizure medications\\
\textbf{EEG:} Electroencephalography\\
\textbf{EMU:} Epilepsy Monitoring Unit\\
\textbf{eTLE:} Extratemporal lobe epilepsy\\
\textbf{FTBTC:} Focal to bilateral tonic clonic (seizure)\\
\textbf{iEEG:} Intracranial EEG\\
\textbf{ILAE:} International League Against Epilepsy\\
\textbf{LSSS:} Liverpool seizure severity scale \citep{baker1991development} \\
\textbf{NHS3:} National Hospital Seizure Severity Scale \citep{o1996national}\\
\textbf{SSQ:} Seizure Severity Questionnaire \citep{cramer2002development}\\
\textbf{TLE:} Temporal lobe epilepsy\\ 
\textbf{Statistical terms}\\
\textbf{AUC:} Area under the curve\\
\textbf{CAR:} Common average reference\\
\textbf{MAD:} Median absolute deviation\\
\textbf{ROC:} Receiver operating characteristic\\

\newpage

\subsection{Patient Metadata\label{suppl:metadata}}
We retrospectively analysed iEEG recordings from a cohort of 63 patients undergoing presurgical evaluation for refractory focal epilepsy. All patients had electrodes surgically implanted as grids and/or strips.
\begin{itemize}
    \item \textbf{Age (yrs):} patient age in years (Median = 29, SD = 7.072).
    \item \textbf{Sex:} patient sex (30 male,  32 female, 1 unknown)
    \item \textbf{Disease duration (yrs):} Time between epilepsy diagnosis and recording in years (Median = 22, SD = 8.581). 
    \item \textbf{Diagnosis:} Purported lobe of onset of the patient’s seizures, based on clinical findings (32 TLE, 25 eTLE, 6 unknown). For this analysis, diagnoses were categorised as TLE or eTLE.
\end{itemize}

The number of subclinical, focal and FTBTC seizures is listed in Supplementary Table~\ref{tab:sz_type_count}; the seizure types experienced by each individual are listed in Supplementary Table~\ref{tab:pat_sz_types}.
\begin{table}[htbp]
    \centering
    \begin{tabular}{lr}
        \hline
        & Number of patients \\
        Seizure type& ($n$=1009)\\
        \hline
        Focal impaired awareness&176\\
        Focal aware&232 \\
        Focal (awareness unknown)&248\\
        Subclinical&323\\
        FTBTC&6\\
        Unknown&24\\
        \hline 
    \end{tabular}
    \caption{Table of counts of seizure types in dataset.}
    \label{tab:sz_type_count}
\end{table}

\begin{table}[htbp]
    \centering
    \begin{tabular}{lr}
    \hline
    & Number of patients \\
    Seizure type&($n$ = 63)\\
    \hline
    Focal only & 29\\
    Subclinical only & 2\\
    FTBTC only & 2\\
    Focal + Subclinical & 23\\
    Focal + FTBTC & 2\\
    Focal + Subclinical + FTBTC	& 2\\
    Unknown	&3\\
    \hline
    \end{tabular}
    \caption{Table of distribution of seizure types for individual patients. }
    \label{tab:pat_sz_types}
\end{table}

\subsection{Supplementary Methods \label{supplmethods}}
\subsubsection{Noise Detection} \label{noisedetect}
Prior to computation of markers and subsequent analysis, each iEEG recording was assessed for noise. Muscle movements and eye blinks were not concerning here as iEEG electrodes are placed directly onto or into the brain and thus are not susceptible to such sources of noise. However, this data was screened for noise from other potential sources. Line noise was removed using a notch filter at 50Hz and 100Hz (with 2Hz windows).  \\
The preictal segment was used to compute a baseline of electrographic activity, which was used in detection of seizure activity and postictal suppression. For more reliable estimates, noise was algorithmically detected as follows:
\begin{enumerate}
    \item Raw iEEG time series MAD scored based on variance and min-max range for each channel independently 
	\item MAD$>$16 labelled as ‘outlier’ - channel is noisy 
	\item ‘Noisy’ channels removed 
	\item iEEG time series common average referenced (CAR)  
	\item MAD$>$16 labelled as ‘outlier’ - channel is noisy 
	\item ‘Noisy’ channels removed 
	\item 1Hz high-pass Butterworth $4^{th}$ order filter used to remove any slow trends 
	\item Repeat the process with a less lenient threshold of MAD$>$12.  
	\item Visual check  
\end{enumerate}
Visual checks were performed to ensure that noise detected was, indeed, noise and to identify potential noise that was not detected. Following this, markers of seizure severity were computed. \\
Noise in the ictal segment was visually assessed using iEEG traces and power spectral density plots - noisy channels were removed from all recordings. We did not seek or remove noise impacting only the postictal segment. 
\subsubsection{Seizure Severity Markers}
We calculated 16 markers of seizure severity based on iEEG recordings. Each marker captures a different aspect of seizure severity; descriptions of the markers and relevant equations are listed in Table~\ref{tab:severity_markers}. 
\begin{table}[htbp]
    \centering
    \begin{tabular}{|l|p{0.2\textwidth}|p{0.35\textwidth} |c|}
         \hline
         Branch&Marker&Description&Equation  \\
         \hline
         \multirow{8}{*}{Peak}& Line length& A measure of the \textbf{complexity} of iEEG signals &\multirow{2}{*}{$\frac{1}{N}\sum^{N-1}_{k=1}|x_{k-1}-x_k|$} \\  \cline{2-4}
         &Energy&\textbf{Relative energy} across all frequency bands& \multirow{2}{*}{$\sum^{N}_{k=1}(x_k-\bar{x})^2$} \\  \cline{2-4}
         & $\delta$ band-power & Power in \textbf{1-4Hz} frequency band&\\ \cline{2-4}
         & $\theta$ band-power&Power in \textbf{4-8Hz} frequency band&\\ \cline{2-4}
         & $\alpha$ band-power&Power in \textbf{8-13Hz} frequency band&\\ \cline{2-4}
         & $\beta$ band-power&Power in \textbf{13-30Hz} frequency band&\\ \cline{2-4}
         & Low $\gamma$ band-power&Power in \textbf{30-60Hz} frequency band&\\ \cline{2-4}
         & High $\gamma$ band-power&Power in\textbf{ 60-100Hz} frequency band&\\ \hline
         \multirow{4}{*}{Spatial}& Proportion of channels included & Proportion of channels with  seizure activity at \textbf{any point} in the ictal period&\\\cline{2-4}
         & Proportion of channels at the point of maximum concurrent activity &Proportion of channels with seizure activity at the point of\textbf{ maximum concurrent} activity &\\ \cline{2-4}
         &Time to max&Time to \textbf{maximum concurrent activity} (in seconds) &\\ \cline{2-4}
         &Prop to max& The \textbf{proportion of seizure duration} to point of maximum concurrent activity&\\ \hline
         \multirow{3}{*}{Suppression}& Majority suppression duration & Time (in seconds) post-ictally with suppression detected in $\geq$ \textbf{80$\%$} of recording channels&\\ \cline{2-4}
          & Partial suppression duration &Time (in seconds) post-ictally with suppression detected in \textbf{$\geq$10$\%$} and \textbf{ $<$80$\%$} of recording channels&\\\cline{2-4}
          & Suppression strength & Median proportion of channels with suppression across the duration of the post-ictal recording &\\ \hline
          N/A&Duration&\textbf{Duration} of the ictal period in seconds (based on visual inspection of iEEG)&\\\hline
    \end{tabular}
    \caption{Table of 16 iEEG-based seizure severity markers proposed within this paper}
    \label{tab:severity_markers}
\end{table}
Our library of objective seizure severity markers has three main branches: peak, spatial, and suppression markers.

\subsubsection{Peak markers} \label{peakmarkers}
Signal complexity was captured using line length\citep{olsen1994automatic}, calculated as:
\begin{center}
    $\frac{1}{N}\sum^{N}_{k=1}|x_{k-1}-x_k|$ 
    \citep{esteller2004comparison}
    \end{center}
The strength of the EEG signal was captured by calculating the signal's energy:
\begin{center}
    $\sum^{N}_{k=1}(x_k-\bar{x})^2$
     \citep{hamad2016feature}
\end{center}
where $\bar{x}$ is the mean of the time series. 

For each severity marker, we first summarised markers across time; for each recording channel, the $95^{th}$ percentile of each marker was calculated. We selected the $95^{th}$ percentile rather than the maximum value to reduce the risk of capturing outlier values which may not have been representative of true seizure activity. The maximum value from this array was then used as the estimated peak activity of the seizure. Each of the peak markers was log-transformed to normalise their distributions. As expected, markers differed across seizure types and patients.

\subsubsection{Spatial markers} \label{spatialmarkers}
Spatial markers were designed to capture the extent of spread of ictal activity across recording channels. Seizure activity was detected using the eight features (line length, energy, band-power in six frequency bands) discussed above. Ictal changes in these features were compared to pre-ictal EEG. For each channel, baseline (pre-ictal) and ictal recordings were split into 1 second, non-overlapping windows. Each of the eight features were calculated for all windows. These computations yielded a baseline distribution of values for each feature and channel. We then scored ictal feature values relative to the baseline distribution to derive if and when a channel was invaded by seizure activity.\\
In detail, the pre-ictal baseline distribution was obtained for each feature and each channel following an automated rejection of pre-ictal spikes or artefacts. We achieved this by removing outliers (in any feature) from the distribution with median absolute deviance (MAD) greater than five. To score each ictal window to the baseline, we used the MAD score, which scores a given observation in terms of the median absolute deviation from the median. We chose MAD scores over $z$-scoring as this method is more robust to outliers.\\
Finally, to derive if any given window in a channel displays seizure activity, we obtained the maximum MAD score across all eight features, effectively measuring if the EEG activity deviated from baseline in any feature. Any window with a maximum MAD score greater than five was deemed as potentially displaying seizure activity. This step yields a binary matrix (of size number of channels by number of time windows) indicating potential seizure activity. To avoid detection of spurious non-seizure activity (e.g. caused by a brief noise or spike), we further validated the binary matrix with a sliding window approach. A symmetric moving sum of length $2 \times \tau +1$ was applied to the binary matrix. If the sum within each sliding window exceeded $\tau$, this window was labelled as having seizure activity. In other words, a channel and time window is deemed to contain seizure activity only if in its temporal vicinity ($\tau$) more than half of windows also showed potential seizure activity. We calculated $\tau$ as 10$\%$ of the seizure duration ($d$). Durations varied from five to 600 seconds in our data; therefore, we bound $\tau$ between two and five seconds to prevent extreme window lengths:

\begin{center}
  $\tau=  \begin{cases}
      2& \text{if}\ d\times0.1<2 \\
      d\times0.1& \text{if}\ 2 \leq d\times 0.1 \leq 5 \\
      5, & \text{if}\ d\times0.1>5\\
    \end{cases}$
\end{center}
In this work, we combined eight markers to capture the spread of seizure activity. For each recording channel, the scale of abnormality compared to the preictal baseline was calculated in each marker. This list of markers is non-exhaustive, it is possible to increase the number of biomarkers included in this algorithm. Future work could expand our list of features (e.g., HFO activity), and we welcome contributions to the library by the community.
\subsubsection{Suppression markers} \label{suppressionmarkers}
Duration and strength of post-ictal suppression was captured by our suppression markers. Signal range was computed as $x_{max}-x_{min}$ in 0.5-second non-overlapping windows. Periods of suppression (calculated in 0.5 second windows) were labelled as majority or partial suppression based on the proportion of suppressed channels: majority suppression was defined as suppression present in over 80$\%$ of recording channels, while partial suppression was defined as suppression between 10$\%$ and 80$\%$ of the channels. Duration of majority and partial suppression were calculated using a 2.5-second moving sum to account for short spikes of activity in suppressed segments. I.e. if a short spike of activity lasted for less than 2.5 seconds, those time points would still be labelled as suppressed. The duration was computed as the time following seizure offset with a one-second buffer. \\
The proportion of seizures with majority suppression differ across seizure types, as reported in Table ~\ref{tab:maj_sup_prop}. As expected, the proportion of seizures with majority suppression increases with the increasing severity of seizure types. Most seizures were found to have post-ictal partial suppression, the only seizures without such suppression had majority suppression for the entire postictal period. The partial suppression marker is likely to suggest suppression in seizures as the threshold for suppression is 5$\%$ of the preictal mean; therefore, with multiple channels and 120-time epochs per channel many instances of suppression will be highlighted by chance.  We invite future work to adjust our threshold of 5$\%$, for example a threshold of 1$\%$ of the baseline will encounter fewer false positives. 
\begin{table}[h!]
    \centering
    \begin{tabular}{cccc}
        \hline
        Seizure type&Subclinical&Focal&FTBTC  \\
         \hline
         Majority suppression&31&172&6\\
         No majority suppression&292&487&0\\
         Proportion of seizures with majority suppression&9.6$\%$&26.1$\%$&100$\%$\\
         \hline
    \end{tabular}
    \caption{Table of counts of seizures with and without majority suppression by seizure type. }
    \label{tab:maj_sup_prop}
\end{table}


\subsubsection{Comparison of marker performance against seizure duration} \label{bootstrap}
As seizure duration is often used to capture severity of seizure severity \citep{beniczky2020biomarkers}, we performed a bootstrapping procedure to compare the across-patient performance of all other markers to seizure duration. We computed a distribution of AUC values based on models created using seizure duration as a marker of seizure severity as follows:
\begin{enumerate}
     \item Create a re-sampled data set by drawing with replacement. For each seizure type in each patient, we re-sampled observations, thereby maintaining the total sample size as well as the number of seizures of each type for each patient.
    \item Created random intercept and random intercept and slope models based on seizure duration, and calculated AUC values for these models.  
    \item Steps 1 and 2 were repeated until 1000 non-Nan AUC values for each model type were computed. The two resultant distributions represented the performance of seizure duration in across-patient analyses.
    \item Calculate the proportion of the duration distribution below the observed AUC for each marker.
\end{enumerate}
 
Each AUC value for the remaining markers were compared the distribution of duration AUC values, the calculated proportion values was used to approximate the scale of improvement beyond the performance of seizure duration. Values were bounded between 0 and 1, with values close to 0 suggesting that duration was the superior marker and values close to 1 suggesting that the alternative marker was superior to seizure duration. The smaller sample size when comparing focal seizures with and without impaired awareness should be noted here, results should be interpreted with caution. 

\subsection{Assessing Marker Performance}
There is currently no gold standard for assessing the severity of epileptic seizures, therefore we validated each of our markers by assessing their performance in classifying seizure types with known differences in severity. Markers were calculated for subclinical (least severe), focal, and FTBTC (most severe) seizures. Markers were validated across all patients and on an individual patient basis. 
\subsubsection{Across patients} \label{acrosspat}
For each model, four hierarchical logistic regression models were created to validate markers. Four models were created for each marker:
\begin{itemize}
    \item\textbf{ Random patient (RP) effects}: Only patient effects are included in the model. This model was used to determine if the distinction between seizure types is driven by patient differences.
    \item \textbf{Fixed marker and random patient effects (random intercept) (RI)}: Here, both fixed marker effects and random patient effects (in the form of random intercepts). This model captures the performance of markers, whilst considering the difference in marker values across patients. 
    \item \textbf{Fixed marker and random patient effects (random intercept and slope) (RIS)}: Here, both fixed marker effects and random patient effects (in the form of random intercepts and slopes). This model captures the performance of markers, whilst considering the difference in marker values and changes in marker values between seizure types across patients. 
\end{itemize}
The performance of each marker was assessed using the area under the curve (AUC) receiver operator curve (ROC). An AUC value of 0.7 or greater was considered acceptable, above 0.8 was considered excellent and above 0.9 was outstanding \citep{mandrekar2010receiver}. Supplementary Table ~\ref{tab:across_focal_subclin} displays the AUC values for each model type in each marker comparing subclinical \textit{vs.} focal seizures. Models with poor fit to the data, as shown by large deviance values, were removed from analysis (AUC is shown here as NaN).

When comparing focal aware and impaired awareness seizures, there were clear patient differences in the marker values; however, the majority of models created with only patient effects were unacceptable classifiers (AUC $<$ 0.7) or poor fit to the data, suggesting that between-patient differences alone did not account for differences between focal aware and impaired awareness seizures. In contrast, 14 severity markers yielded excellent classifier performance with random intercept models or random intercept and slope models. Supplementary Table~\ref{tab:across_aware_unaware} lists AUC values. In validating markers against focal and FTBTC seizure classifications, all markers created excellent or outstanding classifiers using random intercept models. Supplementary Table~\ref{tab:across_focal_ftbtc} lists AUC values. However, the sample size of FTBTC seizures was very small (n=6), therefore the results of this analysis are indicative of good performance but further testing on a larger data set is required. 

\begin{figure}
    \centering
    \includegraphics[width=1\textwidth]{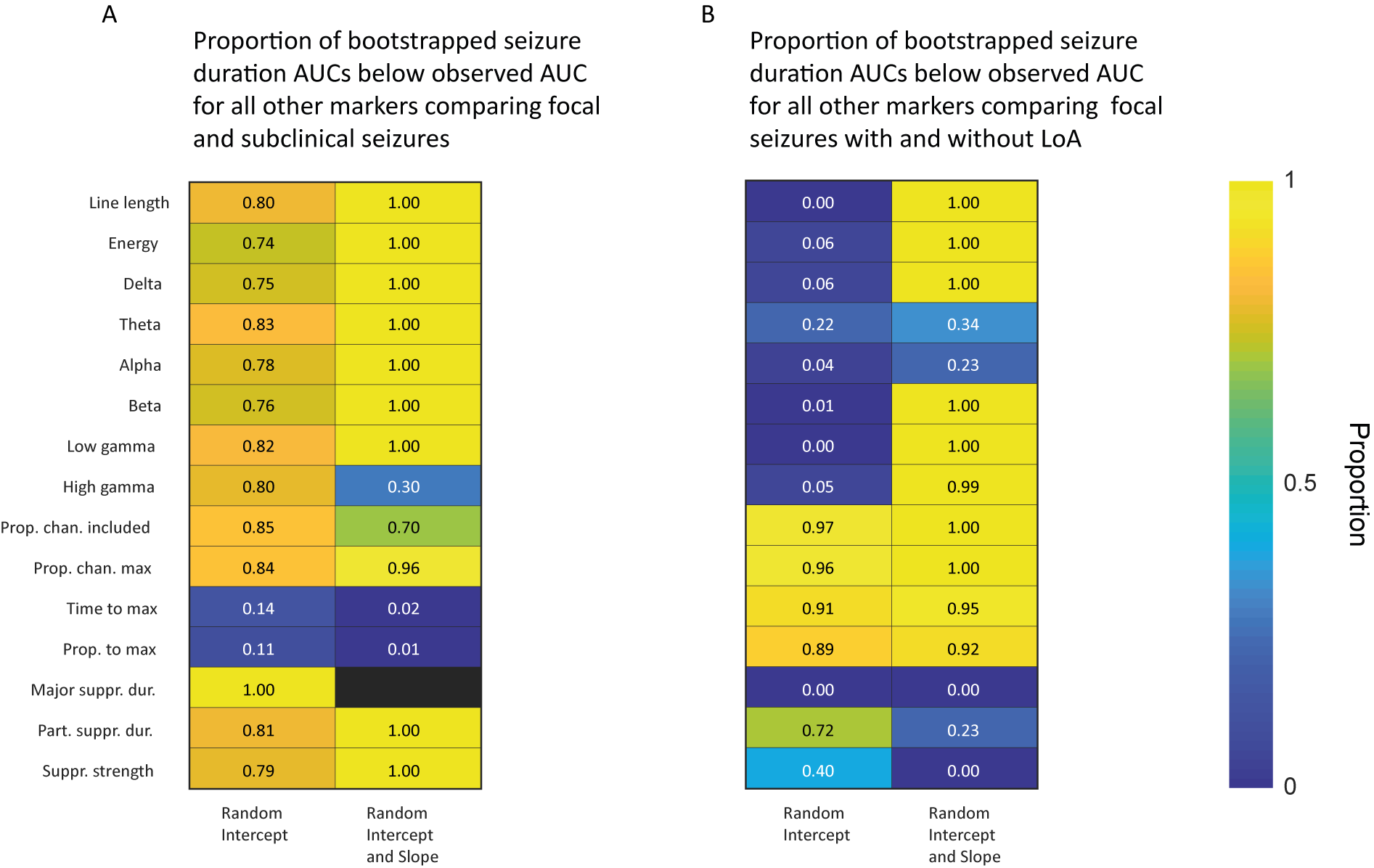}
    \caption{\textbf{Validating markers against ILAE classification across patients.}\\ 
   A) Heat-map representing the proportion of the bootstrapped duration distribution below the observed AUC for all other markers for random intercept and random intercept and slope models comparing focal \textit{vs.} subclinical seizures. B) Heat-map representing the proportion of the bootstrapped duration distribution below the observed AUC for all other markers for random intercept and random intercept and slope models comparing focal seizures with and without impaired awareness. 
   }
    \label{fig:suppl_auc}
\end{figure}

\begin{table}[htbp]
    \centering
    \begin{tabular}{rccc}
    \hline
     \textbf{Marker}&RP&RI&RIS\\
    \hline
    Line length              & NaN   & \textbf{0.934} & 0.881 \\
    Energy                   & NaN   & \textbf{0.926} & \textbf{0.938} \\
    $\delta$ band-power      & NaN   & \textbf{0.928 }& \textbf{0.934} \\
    $\theta$ band-power      & NaN   & \textbf{0.937} & 0.876 \\
    $\alpha$ band-power      & NaN   & \textbf{0.930} & \textbf{0.936} \\
    $\beta$ band-power       & NaN   & \textbf{0.929 }& \textbf{0.923} \\
    Low-$\gamma$ band-power  & NaN   & \textbf{0.935 }& 0.859\\
    High-$\gamma$ band-power & 0.785 & \textbf{0.932} & 0.706 \\
    Prop. chan. included     & 0.800 & \textbf{0.938 }& 0.777 \\
    Prop. chan. at MR        & 0.788 & \textbf{0.937} & 0.823 \\
    Time to MR               & NaN   & \textcolor{gray}{0.635} & \textcolor{gray}{0.620} \\
    Prop. of seizure to MR   & NaN   &  \textcolor{gray}{0.629} & \textcolor{gray}{0.612} \\
    Major. suppr. duration            & 0.668 & \textbf{0.968 }& NaN   \\
    Part. suppr. duration    & NaN   & \textbf{0.935} & \textbf{0.940} \\
    Suppr. strength          & NaN   & \textbf{0.931} & \textbf{0.940 }\\
    Duration                 & NaN   & 0.884 & 0.812 \\
    \hline
\end{tabular}
    \caption{Table of area under the curve (AUC) values for across-patient validation against ILAE classification for subclinical and focal seizures. Outstanding performance is marked in bold. Unacceptable performance is marked in grey.  \\
    \textbf{Shorthand}: RP (model only uses random patient effects), RI (model includes both fixed marker and random patient effects using random intercept), RIS (model includes both fixed marker and random patient effects using random intercept and slope) }
    \label{tab:across_focal_subclin}
\end{table}

\begin{table}[htbp]
    \centering
    \begin{tabular}{rccc}
    \hline
    \textbf{Marker}&RP&RI&RIS\\
        \hline
        Line length & NaN&\textbf{0.917}&\textbf{0.974 }\\
        Energy      & NaN&\textbf{0.952}&\textbf{0.977 }\\
        $\delta$ band-power& NaN&\textbf{0.951} &\textbf{0.970}\\
        $\theta$ band-power & NaN &\textbf{ 0.963 }& 0.866 \\
        $\alpha$ band-power & NaN & \textbf{0.949} & 0.862 \\
        $\beta$ band-power & NaN & \textbf{0.934} &\textbf{ 0.976} \\
        Low-$\gamma$ band-power  & NaN & 0.890 & \textbf{0.970} \\
        High-$\gamma$ band-power & NaN & \textbf{0.950} & \textbf{0.966 }\\
        Prop. chan. included& NaN & \textbf{0.971} & \textbf{0.971 }\\
        Prop. chan. at MR & NaN & \textbf{0.971} & \textbf{0.972} \\
        Time to MR & NaN &\textbf{0.970} & \textbf{0.937} \\
        Prop. of seizure to MR & NaN & \textbf{0.969 }& \textbf{0.935} \\
        Major. suppr. duration & NaN & 0.798 & 0.715 \\
        Part. suppr. duration & NaN & \textbf{0.930 }& 0.850 \\
        Suppr. strength & NaN & \textbf{0.959} & 0.771 \\
        Duration & NaN & \textbf{0.966 }&\textbf{ 0.922}\\
        \hline
    \end{tabular}
    \caption{Table of area under the curve (AUC) values for across-patient validation against ILAE classification for focal aware and impaired awareness seizures. Outstanding performance is marked in bold. \\
    \textbf{Shorthand}: RP (model only uses random patient effects), RI (model includes both fixed marker and random patient effects using random intercept), RIS (model includes both fixed marker and random patient effects using random intercept and slope) }
    \label{tab:across_aware_unaware}
\end{table}

\begin{table}[htbp]
    \centering
    \begin{tabular}{rccc}
    \hline
    \textbf{Marker}&RP&RI&RIS\\
        \hline
    Line length            & 0.844  & \textbf{0.928} & 0.844 \\
    Energy                 & 0.858 & \textbf{0.943 }& 0.777 \\
    $\delta$ band-power    & \textcolor{gray}{0.662} & \textbf{0.962} & \textbf{0.943 }\\
    $\theta$ band-power    & 0.853 & \textbf{0.943 }& 0.818 \\
    $\alpha$ band-power    & 0.800 &\textbf{0.931 }& \textbf{0.957 }\\
    $\alpha$ band-power    & 0.876 & \textbf{0.928 }& 0.842 \\
    Low-$\gamma$           & 0.806 & \textbf{0.943} & \textbf{0.981 }\\
    High $\gamma$          & 0.798 & \textbf{0.971} & 0.750 \\
    Prop. chan. included   & NaN  & \textbf{0.991 }& \textbf{0.991} \\
    Prop. chan. at MR      & NaN & \textbf{0.992} & \textbf{0.992} \\
    Time to MR             & \textcolor{gray}{0.560 }& \textbf{0.977} & \textbf{0.983} \\
    Prop. of seizure to MR & \textcolor{gray}{0.501} &  \textbf{0.977} &\textbf{ 0.984} \\
    Major. suppr. duration          & 0.828  & 0.827 & 0.808 \\
        Part. suppr. duration  & \textcolor{gray}{0.631} & \textbf{0.981} & NaN   \\
    Suppr. strength        & 0.783 & \textbf{0.913} & \textbf{0.992} \\
    Duration               & 0.733 & \textbf{0.972} & \textbf{0.981}\\
    \hline
    \end{tabular}
    \caption{Table of area under the curve (AUC) values for across-patient validation against ILAE classification for focal and FTBTC seizures. It should be noted that only six FTBTC seizures were recorded across all patients. Due to this small sample size, we can only interpret these results as indicative. Outstanding performance is marked in bold. Unacceptable performance is marked in grey. \\
    \textbf{Shorthand}: RP (model only uses random patient effects), RI (model includes both fixed marker and random patient effects using random intercept), RIS (model includes both fixed marker and random patient effects using random intercept and slope) }
    \label{tab:across_focal_ftbtc}
\end{table}

We further divided patients into patients with TLE and eTLE to determine if the performance of markers (focal \textit{vs.} subclinical) was impacted by the lobe in which seizures began. Tables ~\ref{tab:tle_across_focal_subclin} and ~\ref{tab:etle_across_focal_subclin} display AUC values for hierarchical logistic regression models for TLE and eTLE patients, respectively. \\
Comparing AUC values from all patients and TLE and eTLE patients separately, these results qualitatively agree with the results across all patients. Performance of markers across all patients and for TLE or eTLE patients does not differ greatly. It is expected that AUC values are slightly lower as the number of seizures considered is smaller when looking at each diagnosis in turn. These results suggest that markers can be equally applied for TLE and eTLE patients in an across-patient context. It is likely that the difference in markers resulting from different seizure onset zones is captured in the random patient effects. 

\begin{table}[htbp]
    \centering
    \begin{tabular}{rccc}
         \hline
         \textbf{Marker}&RP&RI&RIS\\
        \hline
        Line length              & 0.819 & 0.772 & \textcolor{gray}{0.618} \\
        Energy                   & 0.768  & \textcolor{gray}{0.609} & \textcolor{gray}{0.638} \\
        $\delta$ band-power      & 0.795 & \textcolor{gray}{0.617} & \textcolor{gray}{0.650} \\
        $\theta$ band-power      & \textcolor{gray}{0.688} & 0.785 & \textcolor{gray}{0.500} \\
        $\alpha$ band-power      & NaN  & 0.781 & \textcolor{gray}{0.635} \\
        $\beta$ band-power       & 0.778 & 0.747 & \textcolor{gray}{0.644} \\
        Low-$\gamma$ band-power  & 0.857 & \textcolor{gray}{0.671} & \textcolor{gray}{0.594} \\
        High-$\gamma$ band-power & 0.897 & \textcolor{gray}{0.631} & \textcolor{gray}{0.543} \\
        Prop. chan. included     & \textbf{0.916} & \textcolor{gray}{0.523} & \textcolor{gray}{0.523} \\
        Prop. chan. at MR        & \textbf{0.917 }& \textcolor{gray}{0.503} & \textcolor{gray}{0.502} \\
        Time to MR               & NaN & 0.830 & \textcolor{gray}{0.514} \\
        Prop. of seizure to MR   & NaN   & 0.849 & \textcolor{gray}{0.529} \\
        Major. suppr. duration            & 0.873 & 0.798 & 0.717\\
        Part. suppr. duration    & NaN  & \textbf{0.958} & 0.819 \\
        Suppr. strength          & NaN   & \textbf{0.973} & 0.870 \\
        Duration                 & NaN   & 0.809 & \textcolor{gray}{0.551}\\
    \hline
    \end{tabular}
    \caption{Table of area under the curve (AUC) values for TLE-only across-patient validation against ILAE classification for subclinical and focal seizures. Outstanding performance is marked in bold. Unacceptable performance is marked in grey. \\
    \textbf{Shorthand}: RP (model only uses random patient effects), RI (model includes both fixed marker and random patient effects using random intercept), RIS (model includes both fixed marker and random patient effects using random intercept and slope) }
    \label{tab:tle_across_focal_subclin}
\end{table}

\begin{table}[htbp]
    \centering
    \begin{tabular}{rcccc}
         \hline
         \textbf{Marker}&RP&RI&RIS\\
        \hline
        Line length              & \textcolor{gray}{0.567}  & 0.873 & 0.888 \\
        Energy                   & \textcolor{gray}{0.547}  & 0.871 & 0.899 \\
        $\delta$ band-power      & \textcolor{gray}{0.623}  & 0.873 & 0.896 \\
        $\theta$ band-power      & \textcolor{gray}{0.538} & 0.887 & 0.885 \\
        $\alpha$ band-power      & \textcolor{gray}{0.551}  & 0.859 & 0.883 \\
        $\beta$ band-power       & \textcolor{gray}{0.557}  & 0.889 & \textbf{0.906} \\
        Low-$\gamma$ band-power  & \textcolor{gray}{0.569} & 0.872 & NaN   \\
        High-$\gamma$ band-power & \textcolor{gray}{0.593} & 0.871 & NaN   \\
        Prop. chan. included     & \textcolor{gray}{0.656} & 0.870 & 0.888 \\
        Prop. chan. at MR        & \textcolor{gray}{0.644} & 0.871 & 0.884 \\
        Time to MR               & \textcolor{gray}{0.643} & \textbf{0.904} & \textbf{0.911} \\
        Prop. of seizure to MR   & \textcolor{gray}{0.605} & 0.886 & 0.895 \\
        Major. suppr. duration & \textcolor{gray}{0.630} &  0.879 & 0.881 \\
        Part. suppr. duration    & \textcolor{gray}{0.501} & 0.876 & 0.886 \\
        Suppr. strength          & \textcolor{gray}{0.510} & 0.861 & 0.877 \\
        Duration& 0.734 & \textbf{0.924} & \textbf{0.947}\\
    \hline
    \end{tabular}
    \caption{Table of area under the curve (AUC) values for eTLE-only across-patient validation against ILAE classification for subclinical and focal seizures. Outstanding performance is marked in bold. Unacceptable performance is marked in grey. \\
    \textbf{Shorthand}: RP (model only uses random patient effects), RI (model includes both fixed marker and random patient effects using random intercept), RIS (model includes both fixed marker and random patient effects using random intercept and slope) }
    \label{tab:etle_across_focal_subclin}
\end{table}

\newpage
\subsubsection{Within patients}
Performance of markers in distinguishing focal \textit{vs.} subclinical seizures was assessed using Wilcoxon rank sum tests for each patient. Effect sizes ($r$) were bound between zero and one, with values close to one suggesting strong effects between the seizure types. 
We next investigated potential confounding factors on effect sizes. Here we considered binary variables of patient sex, diagnosis (TLE \textit{vs.} eTLE), and surgery outcome (good \textit{vs.} bad), as well as continuous variables of disease duration, age, number of recording channels, and number of seizures recorded. For binary variables, Wilcoxon rank sum tests were used to assess effects between different patient groups. For continuous variables, we created linear regression models where the effect size is modelled by each variable ($r\sim$ variable). Table ~\ref{tab:com_eff_size} displays $r$ and p-values for all patient variables and each marker .

\begin{table}[htbp]
    \centering
    \begin{tabular}{rccccccc}
        \hline
        \textbf{Marker} & $r_{sex}$ & $r_{TLE}$ &$r_{surgical outcome}$ & $p_{dis.dur.}$ & $p_{age}$ & $p_{n chan}$ & $p_{n sz}$\\
        \hline
    Line length     & 0.085 & 0.276 & 0.000 & 0.716 & 0.525 & 0.515 & 0.060 \\
    Energy          & 0.085 & 0.242 & 0.220 & 0.156 & 0.172 & 0.648 & \textbf{0.042} \\
    $\delta$ band-power & 0.017 & 0.276 & 0.220 & 0.219 & 0.083 & 0.641 & \textbf{0.018 }\\
    $\theta$ band-power & 0.017 & 0.311 & 0.000 & 0.288 & 0.377 & 0.671 & 0.071 \\
    $\alpha$ band-power & 0.222 & 0.069 & 0.110 & 0.662 & 0.321 & 0.581 & 0.097 \\
    $\beta$ band-power  & 0.017 & 0.276 & 0.000 & 0.403 & 0.848 & 0.171 & \textbf{0.020 }\\
    Low-$\gamma$ band-power& 0.222 & 0.345 & 0.220 & 0.947 & 0.968 & 0.911 &0.150 \\
    High-$\gamma$ band-power& 0.085 & 0.207 & 0.275 & 0.476 & 0.257 & 0.989&0.135 \\
    Prop. chan. included& 0.188 & \textbf{0.552} & 0.165 & \textbf{0.002} & 0.571 & \textbf{0.034} &\textbf{ 0.009 }\\
    Prop. chan. at MR   & 0.256 & \textbf{0.552} & 0.055 & \textbf{0.004 }& 0.435 & 0.064 & \textbf{0.006} \\
    Time to MR      & 0.120 & \textbf{0.552} & 0.165 & \textbf{0.006 }& 0.541 & 0.246 & 0.285 \\
    Prop. of seizure to MR.& 0.290 & 0.483 & 0.385 & \textbf{0.002} & 0.460 & 0.260 &0.247 \\
    Major. suppr. duration  & NaN   & NaN   & NaN   & NaN   & NaN   & NaN   & NaN   \\
    Part. suppr. duration & 0.222 & 0.138 & 0.275 & 0.216 & 0.957 & 0.356 & 0.349 \\
    Suppr. strength & 0.085 & 0.069 & 0.055 & 0.903 & 0.421 & 0.841 & 0.202 \\
    Duration    & 0.120 & 0.173 & 0.165 & 0.161 & 0.681 & 0.162 & 0.532 \\
        \hline
        \end{tabular}
     \caption{Comparing Wilcoxon rank sum $r$ values (i.e., effect sizes) across different patient groups. Wilcoxon rank sum was used to compare $r$ values for categorical variables of sex, surgery outcome (ILAE 1 and 2 \textit{vs.} ILAE 3+), and TLE \textit{vs.} eTLE. Resultant $r$ values are presented in first two columns. For continuous variables of disease duration, age, the number of recording channels, and the number of seizures recorded, linear regression with response variable $r$ and continuous patient variables as response variable. Resultant $p$-values for explanatory variables are presented in columns three to six. Effect sizes with $p>0.05$, and $p-$values $>0.05$ are marked in bold. }
    \label{tab:com_eff_size}
\end{table}
These results support that within-patient performance of markers is patient specific, with moderate to large effects between TLE and eTLE patients for six markers. Of the continuous variables, disease duration was found to impact all spatial markers.  The number of recording channels had small effects for proportions of channels included and at maximum recruitment. The number of seizures recorded increased effect sizes for energy, $\delta$, $\beta$, proportion of channels included, and proportion of channels at maximum recruitment. Age did not have any clear effects.

Results suggest that performance of markers is differently impacted by various patient features. This finding supports testing the library of markers on each patient to determine if their performance is adequate for the individual.

\begin{figure}
    \centering
    \includegraphics[width=1\textwidth]{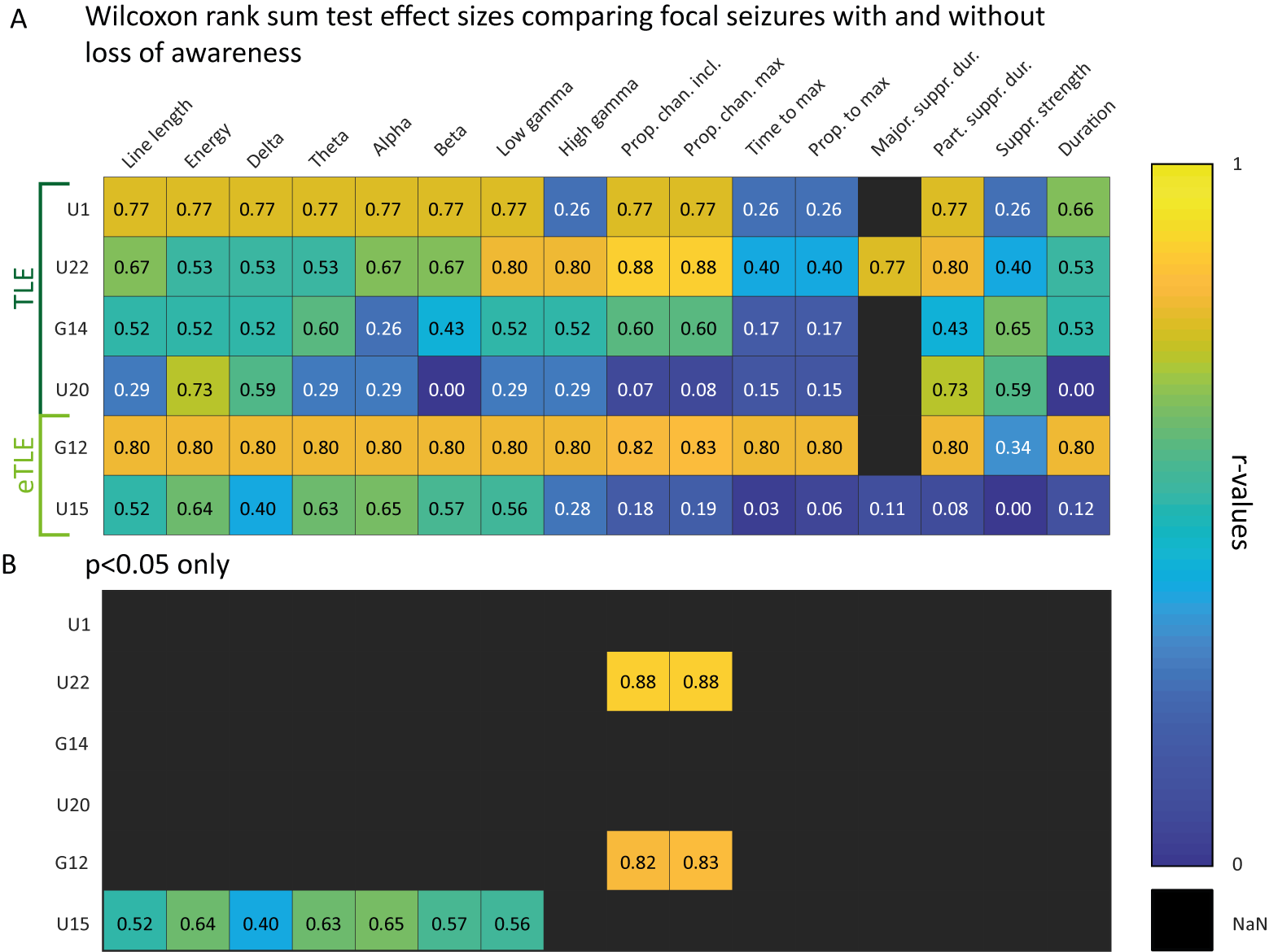}
    \caption{\textbf{Validating markers against focal seizures with and without impaired awareness on a within-patient basis.} A) Heat-map of Wilcoxon Rank Sum $r$ values comparing focal seizures with and without impaired awareness within patients. B) Heat-map of Wilcoxon Rank Sum $r$ values comparing focal seizures with and without impaired awareness within patients with only $r$-values with associated $p$-value $<0.05$. \\
}
    \label{fig:figure_s1}
\end{figure}

Repeating this analysis comparing focal seizures with and without impaired awareness. Fig. ~\ref{fig:figure_s1}A shows a heat-map of $r$-values, Fig.~\ref{fig:figure_s1}B shows $r$-values with associated $p$-value less than 0.05. Only six patients met inclusion criteria for this analysis. In one patient (U15), there are large effects with $p<0.05$ in at least three markers. For patients U22 and G12 two spatial markers (proportion of channels included and at maximum recruitment) had large effect sizes ($r>0.8$, $p<0.05$). Unlike our focal \textit{vs.} subclinical analysis, there is not a clear distinction between TLE and eTLE patients. Further studies with a larger cohort are required to confirm these findings. It was not possible to test differences in effect sizes based on patient metadata as too few patients met inclusion criteria. 

\subsubsection{Capturing fluctuations of seizure severity}
In this paper, we used our markers to capture and assess changes in seizure severity on circadian and longer timescales. Circular-linear correlation was used to assess changes of severity across the day, Table ~\ref{tab:circ_lin_peak} presents circular-linear correlation values for peak markers, all other markers are shown in Table ~\ref{tab:circ_lin_other}. P-values were calculated using permutation tests (1000 permutations). For each permutation, marker values were randomly reassigned, creating `null' models. The proportion of the distribution of `null' test statistics greater than the statistic obtained from the real data was the p-value. Correlations with $p<0.05$ are shown in bold. Note this is reported for reference, and hence no FDR has been applied.
For reference, the probability an individual having at least two or three correlations with $p<0.05$ by chance given 1 marker (i.e., there was no correlation present) was approximately $18.9\%$ or $4.3\%$ respectively given  $\alpha = 0.05$. 
Changes in seizure severity on longer timescales were captured using Spearman's rank correlation between the time of seizure occurrence with respect to the commencement of recording and the marker values. Tables ~\ref{tab:cor_emu_peak} and ~\ref{tab:cor_emu_other} show correlation values for peak and all other markers, respectively. 

\begin{table}[htbp]
    \centering
    \begin{tabular}{rp{1.2cm}p{1.2cm}p{1.2cm}p{1.2cm}p{1.2cm}p{1.2cm}p{1.2cm}p{1.2cm}}
    \hline
    \textbf{Patient ID}& Line length & Energy& $\delta$ band-power&$\theta$ band-power&$\alpha$ band-power&$\beta$ band-power&Low $\gamma $ band-power&High $\gamma$ band-power   \\
    \hline
U9  & \textbf{0.275} & 0.048          & 0.027          & \textbf{0.226} & \textbf{0.358} & \textbf{0.366} & \textbf{0.482} & \textbf{0.342} \\
U13 & 0.050          & 0.033          & 0.094          & 0.037          & 0.017          & 0.091          & 0.095          & 0.062          \\
U14 & 0.070          & \textbf{0.354} & \textbf{0.351} & 0.075          & 0.018          & \textbf{0.129} & 0.026          & 0.068          \\
U15 & \textbf{0.262} & \textbf{0.253} & \textbf{0.161} & \textbf{0.145} & \textbf{0.245} & \textbf{0.320} & \textbf{0.284} & \textbf{0.171} \\
U19 & 0.011          & 0.010          & 0.007          & 0.001          & 0.015          & 0.079          & \textbf{0.123} & 0.071          \\
U22 & \textbf{0.375} & 0.250          & \textbf{0.325} & 0.184          & 0.217          & \textbf{0.338} & \textbf{0.509} & \textbf{0.310} \\
U26 & 0.035          & \textbf{0.370} & 0.234          & 0.204          & 0.008          & 0.040          & \textbf{0.348} & 0.133          \\
U28 & 0.087          & 0.044          & 0.007          & 0.014          & 0.037          & 0.028          & \textbf{0.159} & 0.120          \\
U36 & \textbf{0.159} & 0.021          & 0.045          & \textbf{0.293} & \textbf{0.153} & 0.021          & \textbf{0.099} & 0.040          \\
U43 & 0.053          & \textbf{0.370} & \textbf{0.237} & 0.028          & \textbf{0.251} & 0.145          & 0.062          & 0.055          \\
U46 & 0.023          & 0.033          & 0.052          & 0.036          & 0.002          & 0.050          & 0.009          & 0.015          \\
U48 & 0.073          & 0.042          & 0.085          & 0.022          & 0.034          & 0.005          & 0.003          & 0.135          \\
G4  & \textbf{0.323} & \textbf{0.271} & \textbf{0.234} & \textbf{0.152} & \textbf{0.303} & \textbf{0.259} & \textbf{0.276} & 0.034          \\
G8  & \textbf{0.172} & \textbf{0.079} & \textbf{0.207} & \textbf{0.152} & \textbf{0.087} & \textbf{0.113} & 0.025          & \textbf{0.068} \\
G11 & \textbf{0.134} & 0.003          & 0.004          & 0.020          & 0.010          & \textbf{0.101} & \textbf{0.181} & 0.016         \\
    \hline 
    \end{tabular}
    \caption{Circular linear correlation between markers and time of day of seizure occurrence for peak markers. Correlations with $p<0.05$ based on permutation test with 1000 permutations marked in bold.}
    \label{tab:circ_lin_peak}
\end{table}

\begin{table}[htbp]
    \centering
    \begin{tabular}{rp{1.2cm}p{1.2cm}p{1.2cm}p{1.2cm}p{1.2cm}p{1.2cm}p{1.2cm}p{1.2cm}}
    \hline
    \textbf{Patient ID}&Prop. ch. incl. &Prop. ch. at MR & Time to MR& Prop. to MR & Major. suppr. dur.& Part. suppr. dur. & Suppr. strength &Duration\\
    \hline
U9  & \textbf{0.336} & \textbf{0.399} & 0.009          & 0.012          & 0.125          & 0.146          & 0.191          & 0.019          \\
U13 & 0.084          & 0.064          & 0.006          & 0.043          & NaN            & 0.135          & 0.158          & 0.047          \\
U14 & 0.023          & 0.027          & \textbf{0.281} & \textbf{0.249} & 0.692          & 0.009          & 0.057          & \textbf{0.263} \\
U15 & 0.050          & 0.025          & 0.053          & 0.030          & \textbf{0.423} & 0.096          & 0.016          & 0.124          \\
U19 & 0.073          & 0.040          & 0.014          & 0.019          & 0.006          & 0.051          & 0.097          & 0.005          \\
U22 & \textbf{0.404} & \textbf{0.423} & 0.176          & 0.167          & 0.712          & 0.120          & \textbf{0.304} & 0.083          \\
U26 & 0.171          & 0.139          & 0.080          & 0.078          & NaN             & 0.057          & 0.142          & 0.007          \\
U28 & 0.047          & 0.039          & 0.026          & 0.021          & NaN             & 0.030          & 0.010          & 0.126          \\
U36 & 0.004          & 0.013          & 0.038          & 0.012          & 0.118          & 0.064 & 0.014         & \textbf{0.292} \\
U43 & 0.127          & 0.101          & 0.041          & 0.045          & 0.019          & 0.060          & 0.008          & 0.077          \\
U46 & 0.073          & 0.131          & 0.039          & 0.122          & NaN            & 0.072          & 0.124          & 0.091          \\
U48 & 0.017          & 0.094          & 0.164          & 0.151          & \textbf{0.406} & \textbf{0.373} & \textbf{0.220} & 0.135          \\
G4  & 0.033          & 0.052          & 0.038          & 0.054          & 0.034          & \textbf{0.246} & \textbf{0.210} & 0.002          \\
G8  & 0.012          & 0.041          & \textbf{0.228} & \textbf{0.203} & \textbf{0.313} & \textbf{0.153} & 0.006          & \textbf{0.258} \\
G11 & 0.014          & 0.013          & \textbf{0.075} & \textbf{0.064} & 0.159          & 0.003          & 0.031          & \textbf{0.074}\\
    \hline
    \end{tabular}
    \caption{Circular linear correlation between markers and time of day of seizure occurrence for `spatial` and suppression markers, and duration. Correlations with $p<0.05$ based on permutation test with 1000 permutations marked in bold.}
    \label{tab:circ_lin_other}
\end{table}

\begin{table}[htbp]
    \centering
    \begin{tabular}{rp{1.2cm}p{1.2cm}p{1.2cm}p{1.2cm}p{1.2cm}p{1.2cm}p{1.2cm}p{1.2cm}}
        \hline
    \textbf{Patient ID}& Line length & Energy& $\delta$ band-power&$\theta$ band-power&$\alpha$ band-power&$\beta$ band-power&Low $\gamma $ band-power&High $\gamma$ band-power   \\
    \hline
U9  & 0.320          & 0.223          & 0.084          & 0.303          & 0.212          & \textbf{0.458} & 0.359          & \textbf{0.444} \\
U13 & 0.123          & 0.310          & \textbf{0.487} & 0.211          & 0.095          & 0.048          & 0.015          & 0.004          \\
U14 & \textbf{0.296} & \textbf{0.424} & \textbf{0.426} & \textbf{0.271} & 0.086          & 0.099          & 0.226          & 0.078          \\
U15 & \textbf{0.499} & \textbf{0.653} & \textbf{0.581} & \textbf{0.611} & \textbf{0.576} & \textbf{0.590} & \textbf{0.565} & \textbf{0.289} \\
U19 & \textbf{0.273} & \textbf{0.763} & \textbf{0.759} & \textbf{0.550} & \textbf{0.524} & 0.178          & 0.204          & \textbf{0.479} \\
U22 & 0.278          & \textbf{0.507} & \textbf{0.671} & 0.229          & 0.102          & 0.281          & 0.367          & 0.011          \\
U26 & \textbf{0.442} & \textbf{0.457} & \textbf{0.438} & \textbf{0.483} & 0.408          & \textbf{0.509} & \textbf{0.457} & \textbf{0.479} \\
U28 & 0.073          & \textbf{0.497} & 0.279          & 0.015          & 0.049          & 0.012          & \textbf{0.320} & 0.096          \\
U36 & 0.121          & \textbf{0.510} & \textbf{0.471} & \textbf{0.572} & \textbf{0.404} & \textbf{0.366} & 0.037          & 0.093          \\
U43 & 0.139          & 0.291          & 0.287          & 0.182          & 0.221          & 0.077          & 0.158          & \textbf{0.371} \\
U46 & 0.240          & \textbf{0.757} & \textbf{0.760} & \textbf{0.474} & 0.152          & 0.006          & 0.263          & \textbf{0.466} \\
U48 & 0.211          & 0.025          & 0.327          & 0.037          & 0.283          & 0.357          & 0.254          & 0.122          \\
G4  & 0.116          & 0.023          & 0.056          & 0.079          & 0.055          & 0.151          & \textbf{0.336} & 0.165          \\
G8  & 0.086          & 0.018          & 0.000          & 0.017          & 0.103          & 0.058          & 0.072          & 0.058          \\
G11 & 0.001          & 0.159          & \textbf{0.236} & \textbf{0.228} & 0.060          & 0.152          & 0.118          & \textbf{0.236}\\
    \hline
    \end{tabular}
    \caption{Spearman's rank correlation $\rho$ between peak markers and time in EMU for patients with $\geq 20 $ recorded seizures. Correlations with $p<0.05$ marked in bold}
    \label{tab:cor_emu_peak}
\end{table}

\begin{table}[htbp]
    \centering
    \begin{tabular}{rp{1.2cm}p{1.2cm}p{1.2cm}p{1.2cm}p{1.2cm}p{1.2cm}p{1.2cm}p{1.2cm}}
    \hline
    \textbf{Patient ID}&Prop. ch. incl. &Prop. ch. at MR & Time to MR& Prop. to MR & Major. suppr. dur.& Part. suppr. dur. & Suppr. strength &Duration\\
    \hline
U9  & \textbf{0.574} & \textbf{0.611} & 0.191          & 0.193          & NaN            & 0.230          & 0.207          & 0.080          \\
U13 & \textbf{0.416} & \textbf{0.407} & 0.167          & 0.173          & NaN            & 0.091          & 0.201          & 0.169          \\
U14 & 0.165          & 0.132          & 0.161          & 0.148          & NaN            & 0.015          & 0.032          & 0.213          \\
U15 & 0.049          & 0.011          & 0.084          & 0.041          & NaN            & 0.195          & 0.028          & 0.235          \\
U19 & 0.008          & \textbf{0.383} & 0.239          & \textbf{0.324} & \textbf{0.326} & 0.066          & \textbf{0.503} & 0.871          \\
U22 & 0.356          & 0.407          & 0.257          & 0.239          & NaN            & 0.050          & 0.294          & 0.196          \\
U26 & 0.223          & 0.187          & 0.080          & 0.047          & NaN            & \textbf{0.533} & 0.346          & 0.348          \\
U28 & 0.049          & 0.060          & 0.137          & 0.155          & NaN            & 0.072          & 0.145          & 0.078          \\
U36 & 0.120          & 0.104          & 0.109        & 0.118         & NaN            & \textbf{0.250} & 0.084         & 0.117        \\
U43 & 0.187          & 0.133          & 0.056          & 0.023          & NaN            & 0.176          & 0.024          & 0.291          \\
U46 & 0.340          & 0.356          & 0.366          & 0.320          & NaN            & 0.128          & \textbf{0.429} & 0.233          \\
U48 & 0.238          & \textbf{0.460} & 0.133          & 0.339          & 0.203          & 0.067          & 0.081          & \textbf{0.374} \\
G4  & \textbf{0.264} & 0.229          & \textbf{0.415} & \textbf{0.379} & NaN            & 0.057          & 0.096          & \textbf{0.402} \\
G8  & 0.060          & 0.059          & \textbf{0.279} & \textbf{0.268} & NaN            & 0.066          & 0.149          & \textbf{0.249} \\
G11 & \textbf{0.287} & \textbf{0.279} & \textbf{0.249} & \textbf{0.243} & NaN            & \textbf{0.261} & 0.095          & 0.173   \\
    \hline
    \end{tabular}
    \caption{Absolute Spearman's rank correlation $\rho$ between spatial, suppression $\&$ duration markers, and time in EMU for patients with $\geq 20$ recorded seizures. Correlations with $p<0.05$ marked in bold}
    \label{tab:cor_emu_other}
\end{table}

\end{document}